\documentclass[pdflatex,sn-mathphys-num]{sn-jnl}% Math and Physical Sciences Numbered Reference Style 
\usepackage{graphicx}%
\usepackage{multirow}%
\usepackage{amsmath,amssymb,amsfonts}%
\usepackage{amsthm}%
\usepackage{mathrsfs}%
\usepackage[title]{appendix}%
\usepackage[table]{xcolor}%
\usepackage{adjustbox}%
\usepackage{array}%
\usepackage{caption}%
\usepackage{textcomp}%
\usepackage{manyfoot}%
\usepackage{booktabs}%
\usepackage{algorithm}%
\usepackage{algorithmicx}%
\usepackage{algpseudocode}%
\usepackage{listings}%
\usepackage{parskip}%
\usepackage{graphicx}%%
\usepackage[most]{tcolorbox}%
\usepackage{nameref}%
\usepackage{hyperref}%
\usepackage{setspace}%
\usepackage{multicol}%
\usepackage{mathtools}%
\usepackage{paralist}%
\usepackage{wrapfig}%

\usepackage{xcolor}

\definecolor{background_c}{RGB}{240,240,245}

\AtBeginDocument{%
  \setstretch{1.15}}

\newcommand{\set}[1]{\left\{ #1 \right\} }

\renewcommand{\phi}{\varphi}
\renewcommand{\epsilon}{\varepsilon}

\newcommand{\abs}[1]{\left \arrowvert #1 \right \arrowvert}

\newcommand{\ceil}[1]{\left\lceil #1 \right\rceil}
\newcommand{\floor}[1]{\left\lfloor #1 \right\rfloor}

\newtheorem{giorgio}{$>>>$ WHAT I WOULD LIKE TO SAY}

\theoremstyle{thmstyletwo}%

\theoremstyle{thmstylethree}%


\begin{document}

\title[Article Title]{PoolPy: Automated combinatorial pooling for high-throughput molecular profiling}
%\title[Article Title]{PoolPy: Flexible Combinatorial Group Testing Design for Large-Scale Screening}
%PoolPy: Flexible Implementation and Evaluation of Group Testing Strategies for Large-Scale Screening


\author[1]{\fnm{Lorenzo} \sur{Talamanca}}

\author*[1]{\fnm{Julian} \sur{Trouillon}}\email{jtrouillon@ethz.ch}

\affil[1]{\orgdiv{Institute of Molecular Systems Biology}, \orgname{ETH Zürich}, \orgaddress{\city{Zürich}, \postcode{8093}, \country{Switzerland}}}

%\pagenumbering{gobble}


\abstract{Combinatorial group testing reduces screening costs and turnaround time but remains challenging to apply due to design complexity, varying applicability, and lack of implementation tools. Here we present PoolPy, a unified end-to-end framework and web platform to benchmark, automate and decode combinatorial group testing strategies tailored to application-specific constraints across assay modalities. We demonstrate PoolPy utility for protein-ligand interaction screening and genome-wide molecular profiling, enabling the scaling up of multi-readout functional assays.}


\maketitle

\vspace{-2.8\baselineskip}
{\leftskip=0.06\textwidth \rightskip=0.06\textwidth \keywordfont
\noindent\textbf{Website:} \href{https://trouillon-lab.github.io/PoolPy}{\textcolor{black}{https://trouillon-lab.github.io/PoolPy}}\par
}

\newpage

%\section*{Main}\label{sec1}

Biomedical screening and molecular profiling experiments typically rely on the individual measurement of each sample, which becomes prohibitively expensive and labor-intensive as experimental scales increase.
Combinatorial group testing consists in pooling samples in defined groups before testing to increase the information obtained per test, significantly reducing screening costs and turnaround time \cite{aldridge2019group}. 
This approach is used broadly across fields, including for human infection testing to screen large populations at reduced costs \cite{dorfman1943detection,wein1996pooled,shental2020,mutesa2021pooled}, or for software engineering \cite{xu2022diagnostic,eppstein2007improved}. 
Group testing is also increasingly used in large-scale molecular profiling assays to expand possible search spaces, such as for drug discovery \cite{kainkaryam2009pooling}, CRISPR-based phenotypic screening \cite{Liu2025,Yao2024,Hsiung2025}, or transcriptomics \cite{Cleary2017}, metabolomics \cite{recchia2023}, proteomics \cite{sun2023}, and high-dimensional imaging \cite{Ben-Uri2025} measurements. 

Despite its benefits, combinatorial group testing remains underutilized due to its complex designing and decoding processes, and the lack of accessible design and comparison tools.
In addition, multiple strategies have been proposed with varying performances and ranges of application \cite{du1999combinatorial}, making it challenging to select appropriate designs for specific use cases. 
Due to the lack of implementation tools, studies that use group testing frequently re-implement individual basic methods without comparative evaluation, requiring laborious additional steps and often resulting in suboptimal experimental designs.

%%%%%%%%%%%%
% FIGURE 1
%%%%%%%%%%%%

\begin{figure}[!htbp]
\centering
\begin{adjustbox}{width=1.25\textwidth,center}
  \begin{minipage}{\linewidth}
    \includegraphics[width=\linewidth]{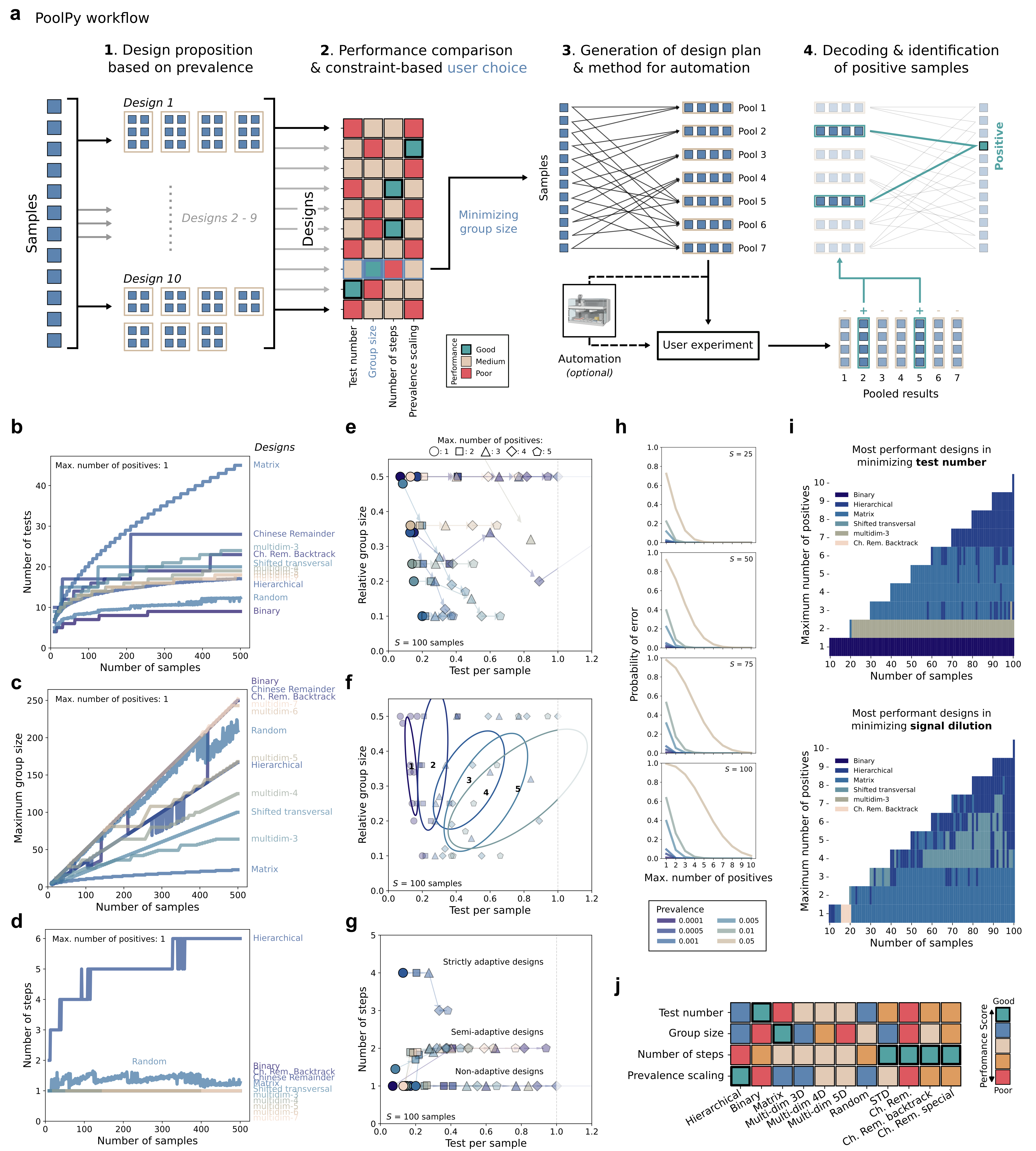}
    \captionsetup{font=footnotesize,singlelinecheck=false}
    \caption{\textbf{Overview of the PoolPy workflow and design performance.} (\textbf{a}) The PoolPy workflow, following four major steps (left to right). (\textbf{b-d}) Comparison of PoolPy designs for number of tests (\textbf{b}), maximum group size (\textbf{c}) and number of steps (\textbf{d}) over 10 to 500 samples for cases with at most one positive sample. (\textbf{e-g}) Relative group size (\textbf{e,f}) and number of steps (\textbf{g}) required by each design related to the number of test per sample over 1 - 5 maximum number of positive samples (reflected by marker shape). Markers are either colored as in \textbf{b} (\textbf{e,g}), or by max. number of positives also indicated with confidence ellipses over one standard deviation (\textbf{f}). (\textbf{h}) Probability of error across prevalence values over 1 - 10 maximum number of positive samples for four total number of samples $S=25,50,75,100$ (top to bottom). (\textbf{i}) Heatmaps showing the best performing PoolPy designs for all combinations of 10 - 100 samples with at most 10\% positive samples in terms of minimizing test number (top) or signal dilution (bottom). (\textbf{j}) Performance summary heatmap of PoolPy designs across four key performance indicators.}
    \label{fig:fig1}
  \end{minipage}
\end{adjustbox}
\end{figure}

%%%%%%%%%%%%
% END FIGURE 1
%%%%%%%%%%%%

Here we present PoolPy, a unified framework and web platform to benchmark, download, automate, and decode combinatorial group testing designs for any application (\hyperref[fig:fig1]{Fig. 1a}; \href{https://trouillon-lab.github.io/PoolPy/}{https://trouillon-lab.github.io/PoolPy/}). 
Unlike existing tools that focus on standard single-readout assays \cite{mclure2021pooltestr,bilder2024bingroup2}, PoolPy provides a complete end-to-end ecosystem, including the generation of method files for robot-assisted automation and the introduction of multi-readout decoding, enabling combinatorial testing for genome-scale molecular profiling assays with complex readouts.
Additionally, while existing tools typically support 1 - 2 methods of interest \cite{mclure2021pooltestr,bilder2024bingroup2}, PoolPy implements 10 conceptually-different algorithms (\hyperref[supp_note_1]{Supplementary Notes 1-2}), including all most used methods and a design introduced here that is based on the theoretical limits of information theory (\hyperref[fig:figSuppBinary]{Fig. S1}), enabling comprehensive benchmarking of design performances. For each use case, PoolPy generates 8 - 13 distinct designs to enable rational selection of optimal designs based on real-world logistical constraints including prevalence (the fraction of expected "positive" hits), cost, time, and signal dilution (\hyperref[supp_note_3]{Supplementary Note 3}). 

To delineate applicability and guide user choice, we benchmarked over $100\small,000$ screening scenarios \textit{in silico} based on key performance indicators (\hyperref[sec_supp]{Dataset S1}; \hyperref[tab:methods]{Table S1}).
At low prevalence, PoolPy generates designs that are highly efficient in reducing test numbers, with as little as nine tests needed to identify up to one positive sample in a set of 500 samples (\hyperref[fig:fig1]{Fig. 1b-d}), or $0.018$ tests per sample. 
To achieve this performance, the corresponding design exploits large groups that contain up to half of all samples (\hyperref[fig:fig1]{Fig. 1c}; \hyperref[fig:figSuppBinary]{Fig. S1}), which can become limiting due to signal dilution (\hyperref[supp_note_3]{Supplementary Note 3}). 
In contrast, designs using smaller groups require increased test numbers (\hyperref[fig:figSuppGroupSizes]{Fig. S2}), reflecting an apparent tradeoff that drives design applicability.

PoolPy shows that designs relying on small group sizes scale better with increased numbers of positive samples (\hyperref[fig:fig1]{Fig. 1e-f}; \hyperref[fig:figSuppGroupSizes]{Fig. S2}), although they then often required an additional step for validations (\hyperref[fig:fig1]{Fig. 1g}; \hyperref[fig:figSuppSteps]{Fig. S3}), or 2 - 6 steps for adaptive designs (\hyperref[fig:figSuppHierarchical]{Fig. S4}). 
In general, group testing performs best at low prevalence, enabling the use of more efficient designs with lower maximum positive numbers (\hyperref[fig:fig1]{Fig. 1f}). 
However, lowering this threshold results in higher error rates in identifying positive samples (\hyperref[fig:fig1]{Fig. 1h}; \hyperref[fig:figSuppPrevalence]{Fig. S5}), highlighting the importance of \textit{a priori} prevalence estimation. 
PoolPy provides guidance and a tool for determining the appropriate pooling hyperparameters to balance this tradeoff based on estimated prevalence (\hyperref[supp_note_4]{Supplementary Note 4}).

Overall, we comprehensively assessed designs performance and showed that no single design emerges as universally optimal (\hyperref[fig:fig1]{Fig. 1i-j}; \hyperref[fig:figSuppMethods]{Fig. S6}), highlighting the need for case-specific design choice based on logistical constraints and screening parameters. 
Specifically, six different PoolPy designs performed best in reducing either test number or signal dilution in at least one scenario across 10 - 100 samples with up to 10\% of positives (\hyperref[fig:fig1]{Fig. 1i}).
Given the broad landscape of group testing applications, PoolPy addresses the current need for a flexible comparative framework to tailor combinatorial designs to specific experimental requirements.

%%%%%%%%%%%%
% FIGURE 2
%%%%%%%%%%%%

\begin{figure}[!htbp]
\centering
\begin{adjustbox}{width=1.25\textwidth,center}
  \begin{minipage}{\linewidth}
    \includegraphics[width=\linewidth]{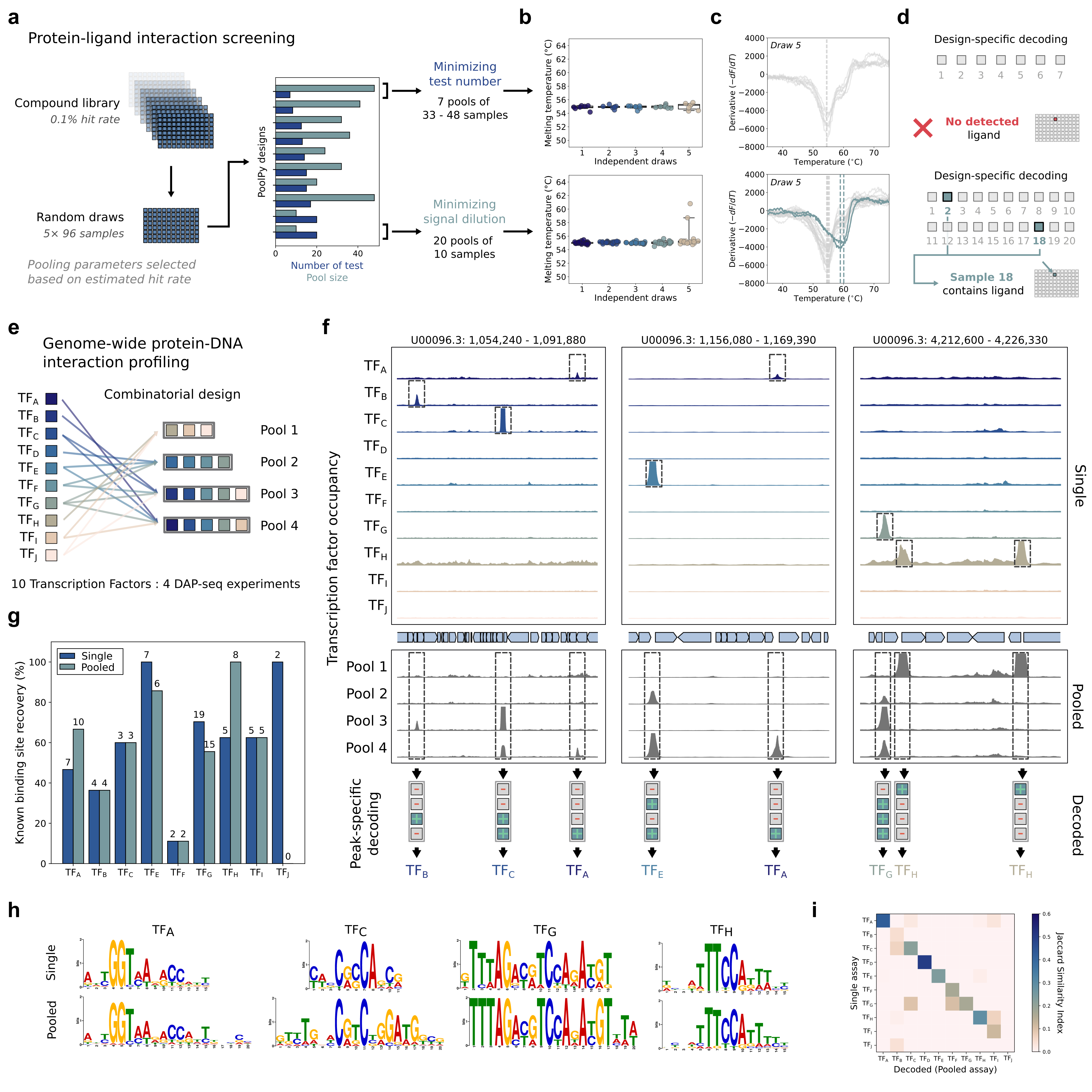}
    \captionsetup{font=footnotesize,singlelinecheck=false}
    \caption{\textbf{PoolPy enables optimal combinatorial pooling across applications.} (\textbf{a}) Schematic representation of protein-ligand interaction screening and the corresponding PoolPy designs. (\textbf{b-d}) Ligand screening for human carbonic anhydrase II across five random sample draws (\textbf{b}), the draw which contained a positive (acetazolamide) sample (\textbf{c}) and the corresponding decoding scheme (\textbf{d}) for designs minimizing test number (top) or signal dilution (bottom). (\textbf{e}) DAP-seq pooling design used in \textbf{f} - \textbf{j} to profile ten \textit{E. coli} TFs in four assays. (\textbf{f}) Transcription factor occupancy (fold enrichment over negative control) tracks over three example regions for single (top) or pooled (bottom) assays. For each region, all 14 plots are scaled to the same y-axis value. (\textbf{g}) Recovery of known binding sites between standard and pooled assays for the nine TFs with annotated binding sites on RegulonDB. (\textbf{h}) Identified DNA motifs in peak regions from single or pooled assays for the four TFs with significant motifs identified. (\textbf{i}) Similarity score matrices between single and pooled assays.}
    \label{fig:fig2}
  \end{minipage}
\end{adjustbox}
\end{figure}

%%%%%%%%%%%%
% END FIGURE 2
%%%%%%%%%%%%

To evaluate PoolPy in an experimental setting, we first performed a protein-ligand interaction screening on a known interaction. 
By testing two distinct PoolPy-generated designs, we investigated the tradeoff between experimental throughput and signal dilution (\hyperref[fig:fig2]{Fig. 2a}). 
Using liquid-handling automation to pool samples from PoolPy-generated automation method files, we screened ligand interactions for the human carbonic anhydrase II enzyme (hCAII) (\hyperref[fig:fig2]{Fig. 2b}). 
The positive samples, containing the known hCAII ligand acetazolamide, were correctly identified among 96 samples by performing 20 pooled tests using the design that minimizes signal dilution, while they were missed by the high-efficiency design using only 7 tests due to higher signal dilution from its larger pools (\hyperref[fig:fig2]{Fig. 2c-d}). 
When increasing ligand concentrations, both designs correctly identified positive samples (\hyperref[fig:figSuppTSA]{Fig. S7}). 
These results demonstrate that the ability of PoolPy to tailor designs to application-specific constraints is essential for robust screening.

Next, we aimed to demonstrate PoolPy use on a high-complexity, multi-readout application by using genome-wide protein-DNA interaction profiling of ten transcription factors (TF) in the model bacterium \textit{Escherichia coli}. 
Using DAP-seq \cite{bartlett2017mapping} to probe genome-wide DNA interactions of the ten TFs, we compared the use of a PoolPy design relying on four pooled tests to the individual profiling of TFs (\hyperref[fig:fig2]{Fig. 2e-f}; \hyperref[sec_supp]{Table S2}). 
Here, we used PoolPy in its multi-readout batch mode, where each genome-wide binding event is considered as an individual readout and is thus uniquely decoded. 
PoolPy assigns each binding event to the corresponding TF according to the combinatorial decoding of the local pooled genomic tracks (\hyperref[fig:fig2]{Fig. 2f}; \hyperref[sec_supp]{Dataset S2}). 
The ten decoded results from the four pooled assays were highly similar to the results obtained with standard individual assays, correctly recovering most known binding sites (\hyperref[fig:fig2]{Fig. 2g}), consensus DNA-binding motifs (\hyperref[fig:fig2]{Fig. 2h}), and expected similarity levels (\hyperref[fig:fig2]{Fig. 2i}). 
Overall, PoolPy correctly identified genome-wide binding sites for ten TFs using four assays, representing a 2.5-fold increase in throughput. 
Such increase would represent a considerable cost reduction for large experiments, whether using DAP-seq on numerous TFs, or for any other compatible profiling assays. If signal dilution is not limiting, cost reductions far greater than 2.5-fold could be achieved (\hyperref[fig:fig1]{Fig. 1b}; \hyperref[fig:figSuppMethods]{Fig. S6}), enabling the exploration of vast search spaces that were previously cost-prohibitive across a diverse range of molecular readouts.
These results establish PoolPy as a robust user-friendly framework for scaling up large-scale functional genomics while maintaining the high resolution of standard individual assays.

In conclusion, PoolPy allows users to compare, apply, automate, and decode optimal combinatorial designs tailored to their specific use cases across any compatible applications, including multi-readout molecular profiling assays.
PoolPy is open-source, includes an accessible web interface and user instructions, and represents a flexible unified framework to integrate potential novel designs in the future.

%%%%%%%%%%%%%%%%%%%%%%%%%%%%%%%%%%%%%%%%%%%%%%%%%%%%%%%%%%%%%%%%%%%%%%%%
% METHODS
\section*{Methods}\label{sec4}

\subsection*{Implementation of group testing designs}
\subsubsection*{General notation}
First, we define $\mathbb{S}=\set{s}$ as the set of $S$ samples with $s=0,...,S-1$.
Of these $S$ samples, we assume that up to $D$ can be positive. 
For the purposes of this description, we consider all positive results to be true positives and all negative results to be true negatives.
The same can be formalized for the $W$ pools, and their set $\mathbb{W}=\set{w}$ with $w=0,...,W-1$.
In general, we represent a set as a math-bold letter, the cardinality as a capital letter, and an element as a lower case letter. 
We use the notation $\mathbb{S}_w$ to indicate the set of samples that are grouped together in pool $w$, as well as $\mathbb{W}_s$ to indicate the set of pools where sample $s$ is present.
$S$ is defined at the beginning of the problem as the number of samples to test, which should be known; 
$W$ is dependent on the pooling strategy used and varies accordingly.
For this reason, $W$ can be written as a function of up to three variables: the method, the number of samples, and the maximum number of positives, $W=W(\text{method},S,D )$.
However, we refer to the number of pools as $W$ to improve readability.
We also use a pool-assigning matrix $PA$, which is a boolean matrix of size $S \times W$ where $(PA)_{sw}=1$ means that sample $s$ is in pool $w$.
According to our previous description and definitions, row $s$ of $PA$ has a simple bijection to $\mathbb{W}_s$ and, conversely, column $w$ has one to $\mathbb{S}_w$.
Defining the $PA$ matrix, $\set{\mathbb{S}_w}$, or $\set{\mathbb{W}_s}$ is equivalent and fully characterizes the pooling strategy.

\subsubsection*{Matrix design}
In the matrix design, all samples are ideally arranged in a square matrix, and all samples in the same row or column are combined into a single pool.
This yields $W \sim 2\sqrt{S}$, which already reduces the experimental burden from $S=6$.
While $S$ cannot always be arranged in a perfect square, this has minimal impact on the effectiveness of this design.
We aim to analytically determine the number of pools needed as well as to express the sets $\mathbb{S}_w$ and, conversely, $\mathbb{W}_s$ in a compact form.
Let the number of rows be $A=\ceil{\sqrt{S}}$ and the number of columns $B=\ceil{S/A}$.
It follows that $A-1\le B\le A$.
To identify the row-derived pools, assuming the matrix is filled left-to-right and top-to-bottom, we have:
\begin{equation}
    \mathbb{S}_a=\set{s|\floor{s/B}=a}.
\end{equation}
Here, the floor operation is equivalent to the computational 'integer division'.
For the column pools, we define:
\begin{equation}
    \mathbb{S}_b=\set{s|s \bmod B = b}.
\end{equation}
Finally, we can define the total number of pools and their assignment as:
\begin{equation}
    W=A+B \quad \mathbb{W}=\mathbb{A}\cup\mathbb{B} \quad w\in \mathbb{W}=
    \begin{cases}
        w\in \mathbb{A} \quad \text{if} \quad w\le A \\
        w-A \in \mathbb{B} \quad \text{if} \quad w\ge A
    \end{cases}
\end{equation}

\subsubsection*{N-dimensional design}
As previously proposed \cite{mutesa2021pooled}, we expand the matrix formalism to any $N$ dimensional matrix design, keeping the equations as general as possible, and assuming that $S\ge2^N$.
The standard matrix formalism does not generalize easily, but we can start by calculating the size of the $N$-dimensional matrix.
Let $L=\ceil{\sqrt[\leftroot{-2}\uproot{2}N]{S}}$ so that $(L-1)^N <S \le L^N $.
Therefore, all sides are of length $L$ or $L-1$, which can be determined iteratively.
Let $L_n$ denote the size of dimension $n$.
There is an arbitrary ordering of all the $L_n$ that we break by defining $\eta$ such that
\begin{equation}
\label{Eq:sim_break_bs}
    L_n=\begin{cases}
        L \quad &n\le\eta\\
        L-1 \quad &n > \eta
    \end{cases}
\end{equation}
By definition, if $\eta =N-1$ then all $L_n=L$.
We now aim to transform a one dimensional number into a set of binary numbers to achieve a pool-assigning function.
We note that each sample belongs to exactly one pool along each dimension.
With this in mind, we can proceed as for the matrix design. When $L_n=L \forall n\in \mathbb{N}| 0\le n\le N-1$, this is equivalent to a base change.
In fact, if all dimensions are of the same size ($L$), passing from a single number identifying the sample to a $N$ dimensional coordinate is equivalent to a base-$L$ expansion.
The sample number $s$ is written in base $L$, padding with zeros to $N$ digits, implicitly defining $\mathbb{W}_s$.
%Each digit then directly indicates the pool for the corresponding dimension.
Formally, we define the map
\begin{equation}
\label{eq:Si_def}
    BL(s)=(s_0, s_1, ..., s_N) \quad | \quad \sum_n L^n \cdot s_n=s
\end{equation}
Each coordinate in this $N$-dimensional space represents which pool of that dimension is the correct one for sample $s$.
By construction we know that each sample is in $N$ distinct pools.
The set of pools for sample $s$, with $s_i$ defined implicitly from Eq. \ref{eq:Si_def} is:
\begin{equation}
\label{Eq:Wi_def}
    \mathbb{W}_s=\set{w|w=nL+s_n \text{ for } n=0,...,N-1}.
\end{equation}
We can conversely define $\mathbb{S}_w$ as 
\begin{equation}
    \mathbb{S}_w=\set{s|s_k=w-Lk}.
\end{equation}
All these definitions require the base transformation of Eq. \ref{eq:Si_def}, which can be performed in parallel, though sequential transformations remain fast for typical sample sizes $< 10^4$.
The above formulas assume that $L_n=L \forall n\in \mathbb{N}|0\le n\le N-1$. When this assumption does not hold, the necessary base-like transformation becomes more complex but can still be similarly defined as:
\begin{equation}
\label{eq:Si_def_gen}
    BN_S(s)=(s_0, s_1, ..., s_N) \quad | \quad \sum_n \left(\prod_{m=0}^n L_m\right) \cdot s_n=s
\end{equation}
where the $BN$ operator generalizes the 'base change' operation to dimension $N$ with $S$ samples.
We remind that having defined $N$ and $S$ uniquely defines all the $L_n$ given Eq. (\ref{Eq:sim_break_bs}).
We can rewrite Eq. \ref{Eq:Wi_def} to fit with the general case:
\begin{equation}
\label{Eq:Wi_def_gen}
    \mathbb{W}_s=\set{w|w=\left(\sum_{m=0}^n L_m\right)+s_n \text{ for } n=0,...,N-1}.
\end{equation}
as well as
\begin{equation}
    \mathbb{S}_w=\set{s|s_k=w-\sum_{m=0}^kL_m}.
\end{equation}
As $(S,N)$ uniquely define $\set{\mathbb{W}_s}$ and $\set{\mathbb{S}_w}$, the $N$ dimensional pooling strategy is fully characterized.

\subsubsection*{Binary design}
\label{sec:binary}
The binary design can be viewed as the maximum dimensional design for every $S$, since it relies on re-writing $s$ in base 2.
In addition, the binary design only measures one out of the two pools of every dimension.
This strategy is highly advantageous when $D=1$, but its performance falls off quickly as soon as $D\ge2$.
We can write as before:
\begin{equation}
\label{eq:S2_def}
    B2(s)=(s_0, s_1, ..., s_{N-1}) \quad | \quad \sum_n 2^n \cdot s_n=s
\end{equation}
The pools are defined as $\mathbb{S}_w=\set{s|s_w=1}$, giving by construction $W=\ceil {\text{log}_2(S)}$.
%This is very similar to high dimensional matrix designs, with the difference that the $0$ pool of each dimension is not considered.
%Doing so allows for a $50\%$ reduction in the number of pools and still yields high information at $d=1$.
%Although ignoring one pool per dimension reduces the total information, for high dimensions this trade-off can be beneficial.
This design truly minimizes the number of tests needed when we are certain that there is exactly one positive. 
However, if we want to consider the case $D=1$, we need to test all samples at least once.
We can do this by leaving empty the $0$th sample such that we can be sure to check for $D=1$ and not only $d=1$.
This adds at most one pool, but in practice often does not, since in many cases $\ceil {\text{log}_2(S)}=\ceil {\text{log}_2(S+1)}$, especially for large $S$.

\subsubsection*{Random design}
The random design consists in setting the number of pools $W$ and the number of samples per pool $S_W$, then extracting for each pool independently $S_W$ objects from $\mathbb{S}$ \cite{bruno1995efficient}.
Once this assignment is done, the design is complete.
In practice, however, using this strategy effectively requires exploring multiple configurations rather than relying on a single draw.
In general, the choice of $N$ and $C$ is not trivial.
Therefore, our implementation strategy has been to test all combinations of $N$ and $C$ such that:
\begin{equation}
    \begin{aligned}
        &mc\le S_W \le MC\\
        &mp \le W \le MP\\
        &mc=\sqrt{S}\\
        &MC=S/2\\
        &mp=\text{log}_2(S)\\
        &MP=2\sqrt{S}.
    \end{aligned}
\end{equation}
Each combination of $N$ and $C$ is tested $5$ times, but to find the real optimal condition many more tests could be needed.
However, we found that fluctuations between different random draws were generally much smaller than the differences observed across configurations, particularly for large $S$.

\subsubsection*{Shifted Transversal design}
For this non-adaptive design previously described \cite{thierry2006new,kainkaryam2008poolhits,xin2009shifted}, it is convenient to consider the transpose of the pool-assigning matrix, so we set $M=(PA)^\intercal$ to follow the notation of the original publication.  
We start by defining a closed rotation of indices, $\sigma$:
\begin{equation}
    \sigma: \mathbb{R}^S \to \mathbb{R}^S \quad \sigma(\bar x)=\sigma\Big((x_0,x_1,..., x_{S-1})\Big)=( x_{S-1}, x_0,x_1,..., x_{S-2}).
\end{equation}
We can apply multiple ($m$) times the $\sigma$ operator, denoted as $\sigma^m$.
For any prime number $q$ such that $q<S$, we define the compressing power of $q$ with respect to $S$, noted $\Gamma(q,S)$ or $\Gamma$, as the smallest integer $\gamma$ for which $q^{\gamma +1}>S$.
The idea behind the shifted transversal design is to construct a simple initial sample-pool assignment and then the full $PA$ matrix by systematically shifting it with the $\sigma$ operator. 
As previously described \cite{thierry2006new}, this design aims to satisfy two conditions: (i) limit the number of pools in which any pair of samples co-occur, and 
(ii) keep the intersection sizes between pools roughly constant, in order to maximize the information content of the design.  
Each layer of the shifted transversal design is composed of $q$ pools, and in each layer every sample is placed in exactly one pool.
Given that we are working with $M$, that is the transpose of $PA$, column $i$ represents the set of pools containing sample $i$.
Let $C_{ij}$ denote the $i$th column in the $j$th layer.  
To break symmetry, in the first ($0$th) layer, the first ($0$th) sample is assigned to the first ($0$th) pool, so that $C_{00}=(1,0,...,0)$. 
The general construction is then given by:
\begin{equation}
    C_{js}=\sigma^{t_{sj}}C_{00} \quad t_{sj}=\sum_{c=0}^\Gamma j^c \floor{\frac{s}{q^c}}.
\end{equation}
This procedure generates, for each $j$, a submatrix with $S$ columns and $q$ rows.  
This method is able to generate $k\le q+1$ layers.
Stacking the first $k$ layers vertically yields the complete pooling matrix $M$ of shape $(q \cdot k) \times S$. 
The construction above does not by itself specify the choice of $q$ or the required number of layers.  
However, necessary conditions can be derived for the method to produce a non-adaptive ($1$-step) pooling design.  
For a given maximum number of positives $D$, one must find a prime $q$ such that
\begin{equation}
    D\cdot \Gamma(q,S) \le q.
\end{equation}
Then, the shifted transversal design is obtained by merging the first $k=D\Gamma(q,S)+1\le q+1$ layers, resulting in a design with $q \cdot k$ pools and guaranteeing a $1$-step pooling experiment.

\subsubsection*{Chinese Remainder design}
This method, based on the Chinese Remainder theorem \cite{eppstein2007improved}, is conceptually similar to the shifted transversal design and also yields a $1$-step pooling strategy. 
Let $\set{p_1^{e_1}, p_2^{e_2}, ...,  p_J^{e_J}}$ be a set of prime numbers $p$ each associated with its exponent $e$ such that
\begin{equation}
\label{eq:constrain_CR}
    \prod_j p_j^{e_j}\ge S^D .
\end{equation}
For each $j$, we construct a pooling submatrix $M_j$, and then stack them vertically into the full pooling matrix $M$.
As above, we take $M = (PA)^\intercal$, so rows correspond to pools and columns to samples.
In this strategy, $M_j$ has size $ t_j\times S$ with $ t_j=p_j^{e_j}$ and therefore $M$ is a $(W=T) \times S$ matrix where
\begin{equation}
   T=\sum_j t_j=\sum_j p_j^{e_j}
\end{equation}
For each $M_j$ we proceed row by row and identify the ones (i.e. which samples are part of that particular pool).
In particular, in matrix $M_j$ we have samples of row $0\le l<t_j$ to be one if and only if column $k$ is equal to $l$ modulo $t_j$:
\begin{equation}
    \left(M_j \right)_{kl}=
    \begin{cases}
        1 \quad l=k \text{ mod }t_j\\
        0 \quad \text{all other cases}
    \end{cases}
\end{equation}
This formula uniquely defines the pooling strategy and provides a constructive strategy to build it once $\set{p_1^{e_1}, p_2^{e_2}, ...,  p_J^{e_J}}$ is know.
In particular, it is proven that this construction yields a $1$-step (non-adaptive) pooling design \cite{eppstein2007improved}.
The simplest variant sets $e_j=1 \space \forall j $ and is generally suboptimal in terms of minimizing $T=W$, but allows for much faster strategy generation and can be sufficient in practice \cite{eppstein2007improved}.

\subsubsection*{Chinese Remainder backtrack}

A different version of the Chinese Remainder method with lower $W$ can be derived by allowing $e_j\neq 1$ and minimizing $T=W$.
A natural question is how to choose the primes and exponents such that $\set{p_1^{e_1}, p_2^{e_2}, ...,  p_J^{e_J}}$ minimizes $T=\sum_j p_j^{e_j}$. The optimal set can be determined using a backtracking approach described here. First, the maximal set of subsequent primes is determined as 
\begin{equation}
    \min_{J} \prod_j^J p_j\ge S^D.
\end{equation} 
Once J is fixed, we can look for the combinations of exponents minimizing $T=W$ respecting Eq. (\ref{eq:constrain_CR}) with the constraint that $p_i^{e_i}\le p_J\space \forall i$.
This faster strategy is implemented in PoolPy as the Backtrack variant of the Chinese Remainder design.

\subsubsection*{Chinese Remainder Special case $D=2$}
The general formalism of the Chinese Remainder construction can be further developed for the special cases of $D=2$ and $D=3$ \cite{eppstein2007improved}, which are based on exploiting different bases. 
For $D=2$, a more efficient pooling strategy (both computationally and in terms of the number of pools) can be obtained by using base-$3$ representations.
In fact, if we define $q$ as $q \vcentcolon = \ceil{\text{log}_3(S)}$, a $1$-step pooling strategy with $W=(q^2+5q)/2$ can be developed.
Following the base change notation adopted in the Binary design section we have:
\begin{equation}
    B3(s)=(s_0,...,s_{q-1}) \quad | \quad \sum_n 3^n \cdot s_n=s.
\end{equation}
Analogously to the binary design, the first $3q$ columns of the pooling matrix $M$ are defined by
\begin{equation}
   \mathbb{S}_w=\set{s|s_{\floor{w/3}}=w \text{ mod } 3}.
\end{equation}
These columns already suffice to identify the position of a single positive sample.
However, in the case of two positives, this construction may be insufficient. 
To fully resolve the $d=2$ case, an additional $\binom{q}{2}$ columns are added.
Let $(q',q'')$ with $0\le q'<q''< q$ denote all $\binom{q}{2}$ possible pairs of natural numbers smaller than $q$.
Following the notation of \cite{eppstein2007improved}, the additional columns of $M$ are then defined as
\begin{equation}
   \mathbb{S}_{q',q''}=\set{s|s_{q'}=s_{q''}}.
\end{equation}
Intuitively, these extra columns can distinguish between two positives by giving a relation of the various digits of the two positives in base-$3$, allowing their unique identification.

\subsubsection*{ Chinese Remainder Special case $D=3$}
The case $D=3$ can also be treated explicitly \cite{eppstein2007improved}.  
Here, a base-$2$ representation is used.
We define $q$ as $q \vcentcolon = \ceil{\text{log}_2(S)}$ to describe a pooling strategy with $W=2q^2-2q$.
Following the notation above we have:
\begin{equation}
    B2(s)=(s_0,...,s_{q-1}) \quad | \quad \sum_n 2^n \cdot s_n=s.
\end{equation}
We again define $(q',q'')$ with $0\le q'<q''< q$ as the $\binom{q}{2}$ possible pairs of natural numbers smaller than $q$. 
We also call $(v',v'')$ the four distinct vectors such that $(v',v'')\in\set{0,1}^2$.
Again, we define the $PA$ matrix as
\begin{equation}
   \mathbb{S}_{q',q'',v',v''}=\set{s|s_{q'}=v', s_{q''}=v''}.
\end{equation}
While the combinatorial argument demonstrating that this construction resolves up to three positives is not straightforward, its correctness has been formally established in \cite{eppstein2007improved}.

\subsubsection*{Hierarchical design}
The hierarchical design is different from all the other designs implemented in PoolPy, as it is the only strictly adaptive method.
This design aims to minimize the number of total experiments by zooming in on positive samples step by step.
The core idea is to split all samples in subsets $\mathbb{S}_w$ (effectively partitioning $\mathbb{S}$) with the properties:
\begin{equation}
\begin{aligned}
    \mathbb{S}_w \cap \mathbb{S}_v=&
    \begin{cases}
    \mathbb{S}_w \quad &\text{if} \quad w=v\\
    \varnothing  \quad &\text{if} \quad w\neq v
    \end{cases}
    \quad = \mathbb{S}_w \delta_{w,v}\\
    &\bigcup_{w\in\mathbb{W}} \mathbb{S}_w =\mathbb{S}\\
    \max_{v,w}&\left(\abs{S_v-S_w}\right)\le 1
\end{aligned}
\end{equation}
where $\delta$ is the Kronecker delta applied to sets.
After this step, we test all pools and restart the procedure iteratively for each positive one.
The case of $D=1$ is of particular theoretical interest.
In fact, a possible intuition would be that the most favorable solution is the binary one: i.e. always splitting the samples into two sets.
In general, we would require:
\begin{equation}
    W=2 \cdot \text{log}_2(S)
\end{equation}
which can be much smaller than $S$, especially for large values.
Ignoring, for simplicity, the integer nature of the problem, we can generalize the above expression assuming that all splits are constant and in $k$ parts at each step. 
This yields:
\begin{equation}
    W_k=k \cdot \text{log}_k(S)=k \cdot \frac{\ln(S)}{\ln(k)}=o(k)
\end{equation}
where we still need to minimize for $k$.
Here, we aim to link this problem with a known one and therefore introduce the function $o(k)$ as the objective function.
The factor $\ln(S)$ does not affect the minimization and is therefore excluded.
As $o(k)>0 \forall k > 1$ is a continuous function, we consider the reciprocal of $o(k)$ and later take the maximum.
We then define the new $o(k)$ as:
\begin{equation}
    o(k)=\frac{\ln(k)}{k}=\ln(k^{1/k})
\end{equation}
We can then safely exponentiate without impacting the value of $k$ for which the maximum occurs, yielding:
\begin{equation}
    o(k)=k^{\frac{1}{k}}.
\end{equation}
The maximum is found when $k=e$.
As only integer splits are feasible, this solution is of limited practical use.
Interestingly, this objective function coincides with the well-known partition problem for maximum product.
In brief, the partition problem for $N$ aims to find the set of numbers such that the sum is equal to $N$ and the product is maximum.
Assuming that all numbers in the set are equal to $n$, their product $P$ is:
\begin{equation}
    P=n^{N/n}=\left(n^{1/n}\right)^N
\end{equation}
Since $N>1$, maximizing $P$ is equivalent to maximizing $n^{1/n}$, which is exactly our objective function.
Since the partition problem has also been solved on integers, the solution is as follows: all $n$ are equal to $3$ apart from one (in the case where $N\bmod 3=2$) or two (in the case where $N\bmod 3=1$) which are equal to $2$.
In the context of pooling, this corresponds to always dividing samples into three pools until we have $4$ or $2$ samples left, when they are divided in two pools.
Here, we defined this strategy for the cases where each pool is tested. In theory, if the number of positives is known, one pool per step can be excluded from testing for further optimization, although this does not reflect typical experimental conditions where the exact number of positives is usually unknown.
For higher $D$, the optimal splitting strategy might differ, and the reported best pooling steps are obtained by exhaustive search.

\subsection*{Pooling parameter determination based on estimated prevalence}
In practice, the precise number of positives in a sample set is typically unknown, but the prevalence in the population can be estimated.
We therefore discuss how to estimate optimal pooling strategies when the true number of positives $\bar d$ is drawn from a binomial probability distribution based on an estimated prevalence value.
It should be noted that when working with prevalence there is always a possibility that a pooling design does not have sufficient resolving power $\big($i.e. $D<\bar d \big)$.
Assuming test results are correct and considering $S$ samples extracted from a population with prevalence $\rho$, for any given $D$ we have:
\begin{equation}
\label{eq:prev_D}
\begin{aligned}
    \mathcal{P}_S\Big(\bar d > D\Big)=\sum_{d=0}^D\rho^d(1-\rho)^{S-d}\binom{S}{d}\simeq \\ 
    \frac{1}{\sqrt{2\pi}\sqrt{S\rho(1-\rho)}} \int_{-\infty}^{D} dx e^{-\frac{(x-D)^2}{2S\rho(1-\rho)}}=\\
    \frac{\sqrt{S}}{\sqrt{2\pi}\sqrt{\rho(1-\rho)}} \int_{-\infty}^{D/S} dx' e^{-\frac{S(x'-D/S)^2}{2\rho(1-\rho)}}.
\end{aligned}
\end{equation}
Eq. (\ref{eq:prev_D}) yields the exact or approximate probability of error when applying a pooling strategy to $S$ samples taken from a population with prevalence $\rho$.
Importantly, it is not necessarily true that the only way to reduce error for fixed $S$ and $\rho$ is to increase $D$.
An alternative is to split the experiment. For instance, instead of pooling all $S$ samples together, one could perform two separate pooling experiments with $S/2$ samples each.
In this case, however, a correction for the Family Wise Error Rate (FWER) is required.
In general:
\begin{equation}
\label{Eq:prob_pre_FWER}
    \mathcal{P}_S\left(\bar d > D\right)=1-\left(\prod_{S_i}\bigg(1- \mathcal{P}_{S_i}\left(\bar d > D_i\right)\bigg) \right) 
\end{equation}
with
\begin{equation}
    \sum_i S_i=S \quad \text{and} \quad \sum_i D_i=D.
\end{equation}
In the case of an equal split into $\eta$ parts, we find that Eq. (\ref{Eq:prob_pre_FWER}) becomes
\begin{equation}
    \mathcal{P}_S\left(\bar d > D\right)=1-\left(1- \mathcal{P}_{S/\eta}\left(\bar d > D/\eta\right) \right)^{\eta}.
\end{equation}
Taking a first-order approximation recovers the usual FWER correction:
\begin{equation}
    \mathcal{P}_S\left(\bar d > D\right)\simeq 1-\left(1- \eta\mathcal{P}_{S/\eta}\left(\bar d > D/\eta\right) \right)=\eta\mathcal{P}_{S/\eta}\left(\bar d > D/\eta\right).
\end{equation}
Finally, we emphasize that the value of $\bar d$ is determined by the two system parameters $S$ and $\rho$, which are either fixed or explicitly specified in the formulas above.

\subsection*{Decoder}
The PoolPy decoder identifies positive samples for a given pooling experiment readout.
Our decoding procedure is optimized for realistic experimental numbers - $10^2-10^4$ samples and up to $1-10$ positives - and might not be optimal in other scenarios.
Let us assume the readout $\mathbb{O}$ to be a binary vector of length $W$.
Let us also define $\mathbb{P}$ as the set of independent combinations that would produce a given readout given a pool assigning matrix and the maximum number of positive samples, $\mathbb{P}=f(\mathbb{O}, PA, D)$.
Ideally, the cardinality of $\mathbb{P}$ would be $P=1$, signaling that the experiment has only one possible solution.
The first step in the decoding procedure is to take only the subset of possible positive samples.
Interpreting $\mathbb{W}_s$ also as a binary vector we can define the quantities:
\begin{equation}
    O_s=\mathbb{O}\cdot \mathbb{W}_s \quad \text{and}\quad W_s=|| \mathbb{W}_s||=\mathbb{W}_s\cdot\mathbb{W}_s
\end{equation}
for each sample $s$.
It is then evident that 
\begin{equation}
    O_s\le W_s \quad \forall s.
\end{equation} 
If $O_s< W_s$ then sample $s$ cannot be positive as there must be at least one well where $s$ is present that is not part of the readout.
Then, we take all possible combinations of samples up to $D$ and test which of them give the correct readout $\mathbb{O}$. In the PoolPy python package, the decoding function tests all possible combinations, while on the web platform the number of possible combinations tested is limited to $10^4$ due to computational limitations. In the majority of cases, and in all cases if the pooling experiment was designed appropriately based on fair prior knowledge, this limit is not reached, resulting in the exact identification of positive samples. If the number of possible combinations exceeds $10^4$, the set of putative possible samples $\set{s|O_s= W_s}$ is returned.

\subsection*{Protein-ligand screen using thermal shift assay}
To simulate ligand screening from a compound library with a prevalence of 0.1\% positive hits, a set of 1,000 samples was generated in 20\% DMSO, in which one sample was randomly spiked with acetazolamide (Sigma, \#A0100000) at either 0.5 or 5 mM. Then, 96 samples were randomly selected from the total set in five independent draws, which were used as testing sets for combinatorial pooling using PoolPy designs. For each draw, two pooling designs were tested on the 96 samples, one minimizing test number and one minimizing signal dilution. In each case, the 96 samples were arranged in a stock 96-well plate and then pooled according to the two designs using a Biomek i7 liquid-handling robot (Beckman Coulter) following the PoolPy-generated automation method file for the corresponding design.

Ligand interaction screening was performed using thermal shift assay with the Protein Thermal Shift Dye Kit (Thermo, \#4461146) in 20 µl reactions containing 10 µM of human carbonic anhydrase II (Sigma, \#C6165), 5 µl of Protein Thermal Shift Buffer, 2.5 µl of Protein Thermal Shift Dye, 2 µl of 10$\times$ PBS and 2 µl of either blank (20\% DMSO) or of pooled samples. Melt curves were then recorded on a QuantStudio 3 Real-Time PCR System (Applied Biosystems) following manufacturer's instructions.

\subsection*{Protein purification}
Protein expression for ten \textit{E. coli} TFs (PdhR, TorR, RutR, DhaR, Mlc, GlpR, MetJ, IclR, UxuR and DgoR; \hyperref[tab:TFTable]{Table S3}) was performed using strains from the ASKA library \cite{kitagawa2005complete}. The expression strains were inoculated at OD\textsubscript{600} = 0.1 in 100 ml of LB medium containing 30 µg/ml chloramphenicol from overnight pre-cultures. After 4 hours of growth at 37°C, protein expression was induced with 0.5 mM of IPTG and cultures were further incubated at 20°C for 20 hours. Cells were then collected by centrifugation and kept at -80°C overnight. Pellets were resuspended in 10 ml of B-PER Complete Bacterial Protein Extraction Reagent (Thermofisher, \#89821) and incubated at room temperature (21°C) for 15 min on a rotating wheel. Lysates were then clarified by centrifugation and the resulting supernatants mixed with 10 ml of IMAC buffer (50 mM Tris-HCl, 500 mM NaCl, 5 \% Glycerol, pH 8) containing 20 mM of imidazole and 500 µl of pre-washed HisPur Cobalt Resin (Thermofisher, \#89964) and further incubated for 1 hour at room temperature on a rotating wheel. Samples were then passed through empty gravity columns (Thermofisher, \#29924) and washed twice with 10 ml of IMAC buffers containing 20 and 40 mM of imidazole each, for a total of four washes. Purified proteins were eluted in 2 ml of IMAC buffer containing 500 mM of imidazole and then dialyzed using dialysis cassettes (Sigma, \#PURX60050) in protein storage buffer (50 mM Tris, 250 mM NaCl, 50 mM KCl, 10 \% Glycerol, 0.5 \% Tween20, pH 7.5). Protein quality and quantity were assessed by SDS-PAGE and using a Qubit protein assay (Thermofisher, \#Q33211).

\subsection*{DAP-seq}
DAP-seq was performed as previously described \cite{trouillon2021transcription} with minor modifications. To generate genomic libraries, genomic DNA was extracted from \textit{E. coli} K-12 MG1655 cells grown overnight in LB medium at 37°C using the Monarch spin gDNA extraction kit (NEB, \#T3010) and then fragmented to 100 - 300 bp by sonication. The genomic DNA was prepared using the NEB End Repair (NEB, \#E6050) and dA Tailing (NEB, \#E6053) modules, and then adaptor-ligated by incubation for 2 hours at room temperature with pre-annealed sequencing Y adaptors (Adaptor A and B; \hyperref[tab:primers]{Table S4}) at a 1:7.5 ratio using the NEB Blunt-TA ligase (NEB, \#E0367), while purifying the resulting DNA after each step using SPRIselect beads (Beckman Coulter, \#B23318) using bead ratios of 1.8, 1.8 and 0.9, respectivelly. Adaptor-ligated libraries were then pre-amplified for 10 cycles using the Phusion High-Fidelity DNA Polymerase (NEB, \#M0530) with 10 ng of template DNA per 50 µl reaction using primers P1 and P2 (\hyperref[tab:primers]{Table S4}) and then cleaned up with 0.9$\times$ SPRIselect beads, yielding final gDNA libraries used in DAP-seq. 

DNA-binding reactions were done in 100 µl of DAP buffer (40 mM Tris-HCl, 150 mM KCl, 10 mM MgCl2, 0.01 \% Triton X-100, pH 7.5) containing 50 ng of adaptor-ligated DNA library, 1 µg of sheared salmon sperm DNA (Thermofisher, \#15632011), 10 µl of pre-washed cobalt magnetic beads (Thermofisher, \#10104D) and 5 µl of 20 µM TF protein sample for TF samples or of protein storage buffer for negative controls. For pooled TF assays, equal volumes of each individual TF proteins were first pooled from 20 µM stocks according to the PoolPy design minimizing test number (four pools for ten TFs). Reactions were incubated for 1 hour at room temperature on a rotating wheel and beads were then washed six times with 200 µl of DAP buffer before resuspension in 25 µl of nuclease-free water and elution by incubation at 95°C for 10 min. The resulting samples were amplified and uniquely barcoded by PCR using the Phusion High-Fidelity DNA Polymerase and primers P5\_universal and P7\_barcoded (\hyperref[tab:primers]{Table S4}). All samples were then pooled, concentrated using 0.9$\times$ SPRIselect beads and cleaned up by gel extraction using the Monarch Spin DNA Gel Extraction Kit (NEB, \#T1120). After quality control and quantification using a 4150 TapeStation (Agilent), the pool was sequenced on a NextSeq2000 (Illumina) using a P2 XLEAP-SBS flowcell (Illumina, \#20100987) in 2 $\times$ 50 bp paired-end mode, yielding an average of 4.8 million reads per sample. DAP-seq reads were mapped to the \textit{E. coli} K-12 U00096.3 genome assembly using Bowtie2 \cite{langmead2012fast} with default parameters. Duplicate fragments were then removed using Samtools \cite{li2009sequence} before peak calling against negative controls using MACS3 \cite{feng2012identifying} in paired-end mode and default parameters, and a minimum fold enrichment of 2. Previously-known binding sites were considered recovered when the entire binding site was within a detected peak region. Peaks shared between samples were identified using a 50\% reciprocal minimum overlap. 
%%%%%%%%%%%%%%%%%%%%%%%%%%%%%%%%%%%%%%%%%%%%%%%%%%%%%%%%%%%%%%%%%%%%%%%%

\backmatter

\section*{Supplementary material}\label{sec_supp}

\textbf{Supplementary Dataset 1: PoolPy precomputed designs and performance comparisons.} This dataset contains designs and performance metrics over a large range of numbers of total samples and positive samples.  

\textbf{Supplementary Dataset 2: DAP-seq peaks.} This dataset contains all identified peaks for single and pooled DAP-seq experiments.  

\textbf{Supplementary Table 1: Performance comparison of group testing methods supported by PoolPy.} This table contains the average performance metrics for each PoolPy designs.  

\textbf{Supplementary Table 2: Decoded DAP-seq peaks.} This table contains all DAP-seq peaks identified in pooled DAP-seq assays with their corresponding decoding results from PoolPy.

\textbf{Supplementary Table 3: List of studied transcription factors.} This table contains the list of the ten \textit{E. coli} TFs used for DAP-seq.  

\textbf{Supplementary Table 4: Primers.} This table contains the list of primers used in this work.

\textbf{Supplementary Figures 1 - 7}: Supporting results. 

\textbf{Supplementary Note 1}: The foundational group testing designs.  

\textbf{Supplementary Note 2}: Methods implemented in PoolPy.  

\textbf{Supplementary Note 3}: The logistical constraints of group testing.  

\textbf{Supplementary Note 4}: PoolPy user guide.    

\section*{Acknowledgments}
This research was funded by the Swiss National Science Foundation (SNSF) Ambizione grant \#PZ00P3\_223880, and by grant \#2023-622 of the Strategic Focus Area “Personalized Health and Related Technologies (PHRT)” of the ETH Domain (Swiss Federal Institutes of Technology). The sequencing was performed using instruments from the Functional Genomics Center Zurich (FGCZ) of University of Zurich and ETH Zurich.

\section*{Data and code availability}
The PoolPy web application and tools are available at \href{https://trouillon-lab.github.io/PoolPy}{https://trouillon-lab.github.io/PoolPy}. 
The PoolPy code used for this study is available through GitHub at \href{https://github.com/trouillon-lab/PoolPy}{https://github.com/trouillon-lab/PoolPy}.
Supplementary Dataset 1 containing precomputed designs and performance comparisons is available at \href{https://doi.org/10.5281/zenodo.18660062}{https://doi.org/10.5281/zenodo.18660062}. 
%The raw DAP-seq sequencing data is available in the NCBI database under the bioproject number \href{https://www.ncbi.nlm.nih.gov/bioproject/PRJNA1425967}{PRJNA1425967}.

\bibliography{sn-bibliography}% common bib file

@article{kainkaryam2009pooling,
  title={Pooling in high-throughput drug screening},
  author={Kainkaryam, Raghunandan M and Woolf, Peter J},
  journal={Current opinion in drug discovery \& development},
  volume={12},
  number={3},
  pages={339},
  year={2009},
  publisher={NIH Public Access}
}

@article{thierry2006new,
  title={A new pooling strategy for high-throughput screening: the Shifted Transversal Design},
  author={Thierry-Mieg, Nicolas},
  journal={BMC bioinformatics},
  volume={7},
  pages={1--13},
  year={2006},
  publisher={Springer}
}

@article{aldridge2019group,
  title={Group testing: an information theory perspective},
  author={Aldridge, Matthew and Johnson, Oliver and Scarlett, Jonathan and others},
  journal={Foundations and Trends in Communications and Information Theory},
  volume={15},
  number={3-4},
  pages={196--392},
  year={2019},
  publisher={Now Publishers, Inc.}
}

@article{mutesa2021pooled,
  title={A pooled testing strategy for identifying SARS-CoV-2 at low prevalence},
  author={Mutesa, Leon and Ndishimye, Pacifique and Butera, Yvan and Souopgui, Jacob and Uwineza, Annette and Rutayisire, Robert and Ndoricimpaye, Ella Larissa and Musoni, Emile and Rujeni, Nadine and Nyatanyi, Thierry and others},
  journal={Nature},
  volume={589},
  number={7841},
  pages={276--280},
  year={2021},
  publisher={Nature Publishing Group UK London}
}

@article{bruno1995efficient,
  title={Efficient pooling designs for library screening},
  author={Bruno, William J and Knill, Emanuel and Balding, David J and Bruce, David C and Doggett, Norman A and Sawhill, Wesley W and Stallings, Raymond L and Whittaker, Craig C and Torney, David C},
  journal={Genomics},
  volume={26},
  number={1},
  pages={21--30},
  year={1995},
  publisher={Elsevier}
}

@article{kainkaryam2008poolhits,
  title={poolHiTS: A Shifted Transversal Design based pooling strategy for high-throughput drug screening},
  author={Kainkaryam, Raghunandan M and Woolf, Peter J},
  journal={BMC bioinformatics},
  volume={9},
  pages={1--11},
  year={2008},
  publisher={Springer}
}

@article{xin2009shifted,
  title={Shifted Transversal Design smart-pooling for high coverage interactome mapping},
  author={Xin, Xiaofeng and Rual, Jean-Fran{\c{c}}ois and Hirozane-Kishikawa, Tomoko and Hill, David E and Vidal, Marc and Boone, Charles and Thierry-Mieg, Nicolas},
  journal={Genome research},
  volume={19},
  number={7},
  pages={1262--1269},
  year={2009},
  publisher={Cold Spring Harbor Lab}
}

@article{du1999combinatorial,
  title={Combinatorial group testing and its applications},
  author={Du, Ding-Zhu and Hwang, Frank Kwang Ming},
  volume={12},
  year={1999},
  journal={World Scientific}
}

@article{eppstein2007improved,
author = { Eppstein, David and  Goodrich, Michael T. and  Hirschberg, Daniel S.},
title = {Improved Combinatorial Group Testing Algorithms for Real‐World Problem Sizes},
journal = {SIAM Journal on Computing},
volume = {36},
number = {5},
pages = {1360-1375},
year = {2007},
}

@article{hou2017hierarchical,
  title={Hierarchical group testing for multiple infections},
  author={Hou, Peijie and Tebbs, Joshua M and Bilder, Christopher R and McMahan, Christopher S},
  journal={Biometrics},
  volume={73},
  number={2},
  pages={656--665},
  year={2017},
  publisher={Wiley Online Library}
}

@article{dorfman1943detection,
  title={The detection of defective members of large populations},
  author={Dorfman, Robert},
  journal={The Annals of mathematical statistics},
  volume={14},
  number={4},
  pages={436--440},
  year={1943},
  publisher={JSTOR}
}

@article{wein1996pooled,
  title={Pooled testing for HIV screening: capturing the dilution effect},
  author={Wein, Lawrence M and Zenios, Stefanos A},
  journal={Operations Research},
  volume={44},
  number={4},
  pages={543--569},
  year={1996},
  publisher={INFORMS}
}

@article{xu2022diagnostic,
  title={The diagnostic accuracy of pooled testing from multiple individuals for the detection of Chlamydia trachomatis and Neisseria gonorrhoeae: a systematic review},
  author={Xu, Yangqi and Aboud, Lily and Chow, Eric PF and Mello, Maeve B and Wi, Teodora and Baggaley, Rachel and Fairley, Christopher K and Peeling, Rosanna and Ong, Jason J},
  journal={International Journal of Infectious Diseases},
  volume={118},
  pages={183--193},
  year={2022},
  publisher={Elsevier}
}

@article{barillot1991,
    author = {Barillot, Emmanuel and Lacroix, Bruno and Cohen, Daniel},
    title = {Theoretical analysis of library screening using a N-dimensional pooling strategy},
    journal = {Nucleic Acids Research},
    volume = {19},
    number = {22},
    pages = {6241-6247},
    year = {1991},
    month = {11},
    issn = {0305-1048}
}

@article{shental2020,
author = {Noam Shental  and Shlomia Levy  and Vered Wuvshet  and Shosh Skorniakov  and Bar Shalem  and Aner Ottolenghi  and Yariv Greenshpan  and Rachel Steinberg  and Avishay Edri  and Roni Gillis  and Michal Goldhirsh  and Khen Moscovici  and Sinai Sachren  and Lilach M. Friedman  and Lior Nesher  and Yonat Shemer-Avni  and Angel Porgador  and Tomer Hertz },
title = {Efficient high-throughput SARS-CoV-2 testing to detect asymptomatic carriers},
journal = {Science Advances},
volume = {6},
number = {37},
pages = {eabc5961},
year = {2020}}

@Article{Liu2025,
author={Liu, Nuo
and Kattan, Walaa E.
and Mead, Benjamin E.
and Kummerlowe, Conner
and Cheng, Thomas
and Ingabire, Sarah
and Cheah, Jaime H.
and Soule, Christian K.
and Vrcic, Anita
and McIninch, Jane K.
and Triana, Sergio
and Guzman, Manuel
and Dao, Tyler T.
and Peters, Joshua M.
and Lowder, Kristen E.
and Crawford, Lorin
and Amini, Ava P.
and Blainey, Paul C.
and Hahn, William C.
and Cleary, Brian
and Bryson, Bryan
and Winter, Peter S.
and Raghavan, Srivatsan
and Shalek, Alex K.},
title={Scalable, compressed phenotypic screening using pooled perturbations},
journal={Nature Biotechnology},
year={2025},
month={Aug},
day={01},
volume={43},
number={8},
pages={1324-1336},
issn={1546-1696}
}

@Article{Yao2024,
author={Yao, Douglas
and Binan, Loic
and Bezney, Jon
and Simonton, Brooke
and Freedman, Jahanara
and Frangieh, Chris J.
and Dey, Kushal
and Geiger-Schuller, Kathryn
and Eraslan, Basak
and Gusev, Alexander
and Regev, Aviv
and Cleary, Brian},
title={Scalable genetic screening for regulatory circuits using compressed Perturb-seq},
journal={Nature Biotechnology},
year={2024},
month={Aug},
day={01},
volume={42},
number={8},
pages={1282-1295},
issn={1546-1696}
}

@Article{Cleary2017,
author={Cleary, Brian
and Cong, Le
and Cheung, Anthea
and Lander, Eric S.
and Regev, Aviv},
title={Efficient Generation of Transcriptomic Profiles by Random Composite Measurements},
journal={Cell},
year={2017},
month={Nov},
day={30},
publisher={Elsevier},
volume={171},
number={6},
pages={1424-1436.e18},
issn={0092-8674}
}

@article{recchia2023,
author = {Recchia, Michael J. J. and Baumeister, Tim U. H. and Liu, Dennis Y. and Linington, Roger G.},
title = {MultiplexMS: A Mass Spectrometry-Based Multiplexing Strategy for Ultra-High-Throughput Analysis of Complex Mixtures},
journal = {Analytical Chemistry},
volume = {95},
number = {32},
pages = {11908-11917},
year = {2023},
note ={PMID: 37530514}
}

@article{sun2023,
author = {Sun, Huan and Yang, Ka and Zhang, Xue and Fu, Yingxue and Yarbro, Jay and Wu, Zhiping and Chen, Ping-Chung and Chen, Taosheng and Peng, Junmin},
title = {Evaluation of a Pooling Chemoproteomics Strategy with an FDA-Approved Drug Library},
journal = {Biochemistry},
volume = {62},
number = {3},
pages = {624-632},
year = {2023},
note ={PMID: 35969671}
}

@Article{Ben-Uri2025,
author={Ben-Uri, Raz
and Ben Shabat, Lior
and Shainshein, Dana
and Bar-Tal, Omer
and Bussi, Yuval
and Maimon, Noa
and Keidar Haran, Tal
and Milo, Idan
and Goliand, Inna
and Addadi, Yoseph
and Salame, Tomer Meir
and Rochwarger, Alexander
and Sch{\"u}rch, Christian M.
and Bagon, Shai
and Elhanani, Ofer
and Keren, Leeat},
title={High-dimensional imaging using combinatorial channel multiplexing and deep learning},
journal={Nature Biotechnology},
year={2025},
month={Mar},
day={25},
issn={1546-1696}
}

@Article{Hsiung2025,
author={Hsiung, C. C.-S.
and Wilson, C. M.
and Sambold, N. A.
and Dai, R.
and Chen, Q.
and Teyssier, N.
and Misiukiewicz, S.
and Arab, A.
and O'Loughlin, T.
and Cofsky, J. C.
and Shi, J.
and Gilbert, L. A.},
title={Engineered CRISPR-Cas12a for higher-order combinatorial chromatin perturbations},
journal={Nature Biotechnology},
year={2025},
month={Mar},
day={01},
volume={43},
number={3},
pages={369-383},
issn={1546-1696}
}

@article{kitagawa2005complete,
  title={Complete set of ORF clones of Escherichia coli ASKA library (A Complete S et of E. coli K-12 ORF A rchive): Unique Resources for Biological Research},
  author={Kitagawa, Masanari and Ara, Takeshi and Arifuzzaman, Mohammad and Ioka-Nakamichi, Tomoko and Inamoto, Eiji and Toyonaga, Hiromi and Mori, Hirotada},
  journal={DNA research},
  volume={12},
  number={5},
  pages={291--299},
  year={2005},
  publisher={Oxford University Press}
}

@article{trouillon2021transcription,
  title={Transcription inhibitors with XRE DNA-binding and cupin signal-sensing domains drive metabolic diversification in Pseudomonas},
  author={Trouillon, Julian and Ragno, Michel and Simon, Victor and Attr{\'e}e, Ina and Elsen, Sylvie},
  journal={Msystems},
  volume={6},
  number={1},
  pages={10--1128},
  year={2021},
  publisher={American Society for Microbiology 1752 N St., NW, Washington, DC}
}

@article{langmead2012fast,
  title={Fast gapped-read alignment with Bowtie 2},
  author={Langmead, Ben and Salzberg, Steven L},
  journal={Nature methods},
  volume={9},
  number={4},
  pages={357--359},
  year={2012},
  publisher={Nature Publishing Group}
}

@article{li2009sequence,
  title={The sequence alignment/map format and SAMtools},
  author={Li, Heng and Handsaker, Bob and Wysoker, Alec and Fennell, Tim and Ruan, Jue and Homer, Nils and Marth, Gabor and Abecasis, Goncalo and Durbin, Richard and 1000 Genome Project Data Processing Subgroup},
  journal={bioinformatics},
  volume={25},
  number={16},
  pages={2078--2079},
  year={2009},
  publisher={Oxford University Press}
}

@article{feng2012identifying,
  title={Identifying ChIP-seq enrichment using MACS},
  author={Feng, Jianxing and Liu, Tao and Qin, Bo and Zhang, Yong and Liu, Xiaole Shirley},
  journal={Nature protocols},
  volume={7},
  number={9},
  pages={1728--1740},
  year={2012},
  publisher={Nature Publishing Group UK London}
}

@article{bartlett2017mapping,
  title={Mapping genome-wide transcription-factor binding sites using DAP-seq},
  author={Bartlett, Anna and O'Malley, Ronan C and Huang, Shao-shan Carol and Galli, Mary and Nery, Joseph R and Gallavotti, Andrea and Ecker, Joseph R},
  journal={Nature protocols},
  volume={12},
  number={8},
  pages={1659--1672},
  year={2017},
  publisher={Nature Publishing Group UK London}
}

@article{mclure2021pooltestr,
  title={PoolTestR: An R package for estimating prevalence and regression modelling for molecular xenomonitoring and other applications with pooled samples},
  author={McLure, Angus and O'Neill, Ben and Mayfield, Helen and Lau, Colleen and McPherson, Brady},
  journal={Environmental Modelling \& Software},
  volume={145},
  pages={105158},
  year={2021},
  publisher={Elsevier}
}

@article{bilder2024bingroup2,
  title={binGroup2: statistical tools for infection identification via group testing},
  author={Bilder, Christopher R and Hitt, Brianna D and Biggerstaff, Brad J and Tebbs, Joshua M and McMahan, Christopher S},
  journal={The R journal},
  volume={15},
  number={4},
  pages={21},
  year={2024}
}

\newpage

%%%%%%%%%%%%
% SUPPLEMENTARY FIGURES

\renewcommand{\thefigure}{S\arabic{figure}}
\setcounter{figure}{0}

%\pagenumbering{gobble}

%%%%%%%%%%%%%%%%%%%%%%%%%%%%%%%%%%%%%%%%%%%%%%%%%%%%%%%%%%%%%%%%%%%%%%%%
% FIGURE S1
\begin{figure}[!t]
\centering
\begin{adjustbox}{width=1.35\textwidth,center}
  \begin{minipage}{\linewidth}
    \includegraphics[width=\linewidth]{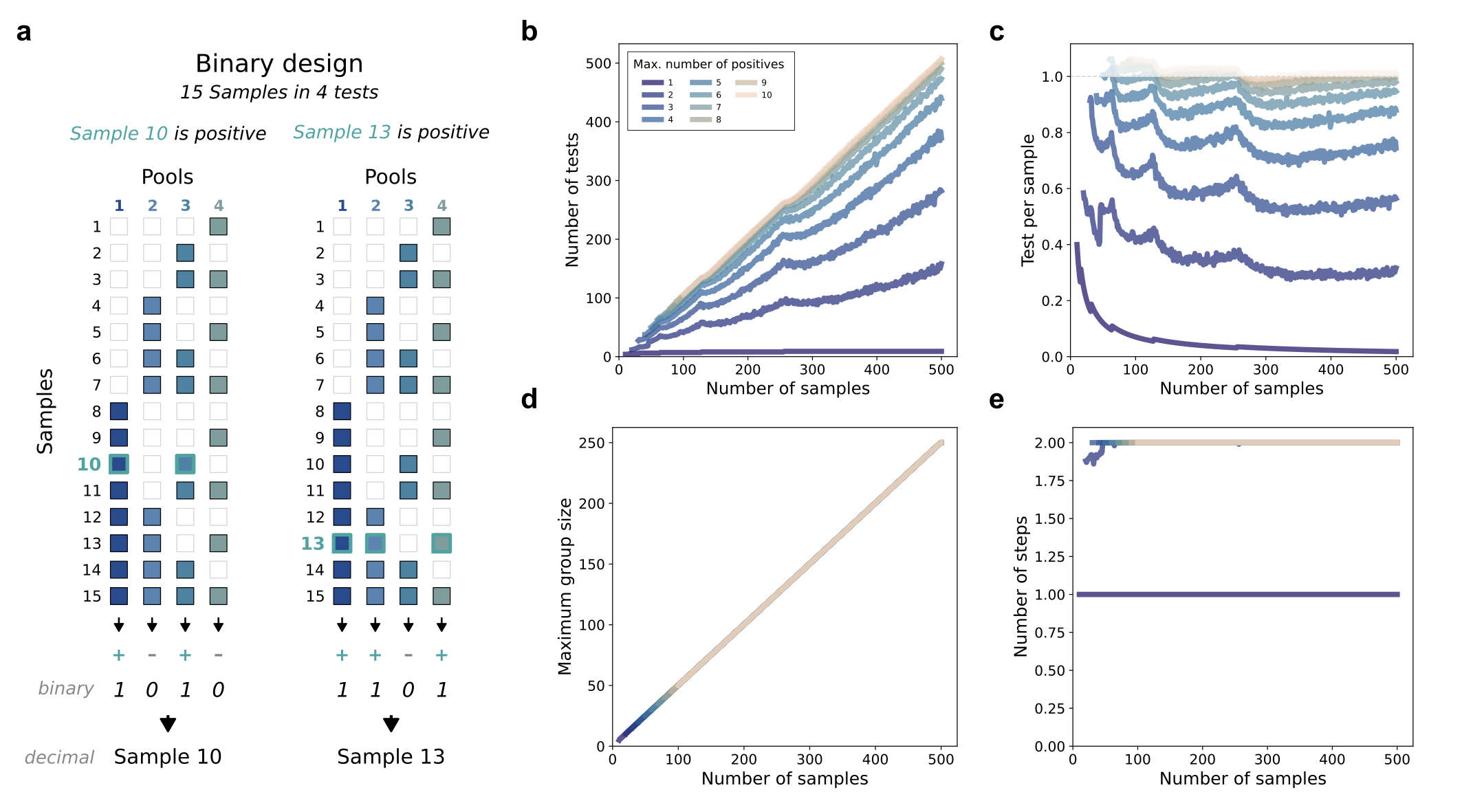}
    \captionsetup{font=footnotesize,singlelinecheck=false}
    \caption{\textbf{The binary design performance decreases sharply with increasing prevalence.} (\textbf{a}) Schematic illustration of the binary design. Two examples are shown with each a different positive sample out of 15. For 15 samples, the binary design makes four pools of eight samples each. The result pattern of the four pools encodes the identity of the positive sample in binary numeral system. (\textbf{b-e}) Number of total tests (\textbf{b}), number of test per sample (\textbf{c}), maximum group size (\textbf{d}) or number of steps (\textbf{e}) needed using the binary design with 1 - 10 maximum numbers of positive samples across 10 to 100 samples.}
    \label{fig:figSuppBinary}
  \end{minipage}
\end{adjustbox}
\end{figure}
\clearpage

%%%%%%%%%%%%%%%%%%%%%%%%%%%%%%%%%%%%%%%%%%%%%%%%%%%%%%%%%%%%%%%%%%%%%%%%
% FIGURE S2
\begin{figure}[!t]
\centering
\begin{adjustbox}{width=1.35\textwidth,center}
  \begin{minipage}{\linewidth}
    \includegraphics[width=\linewidth]{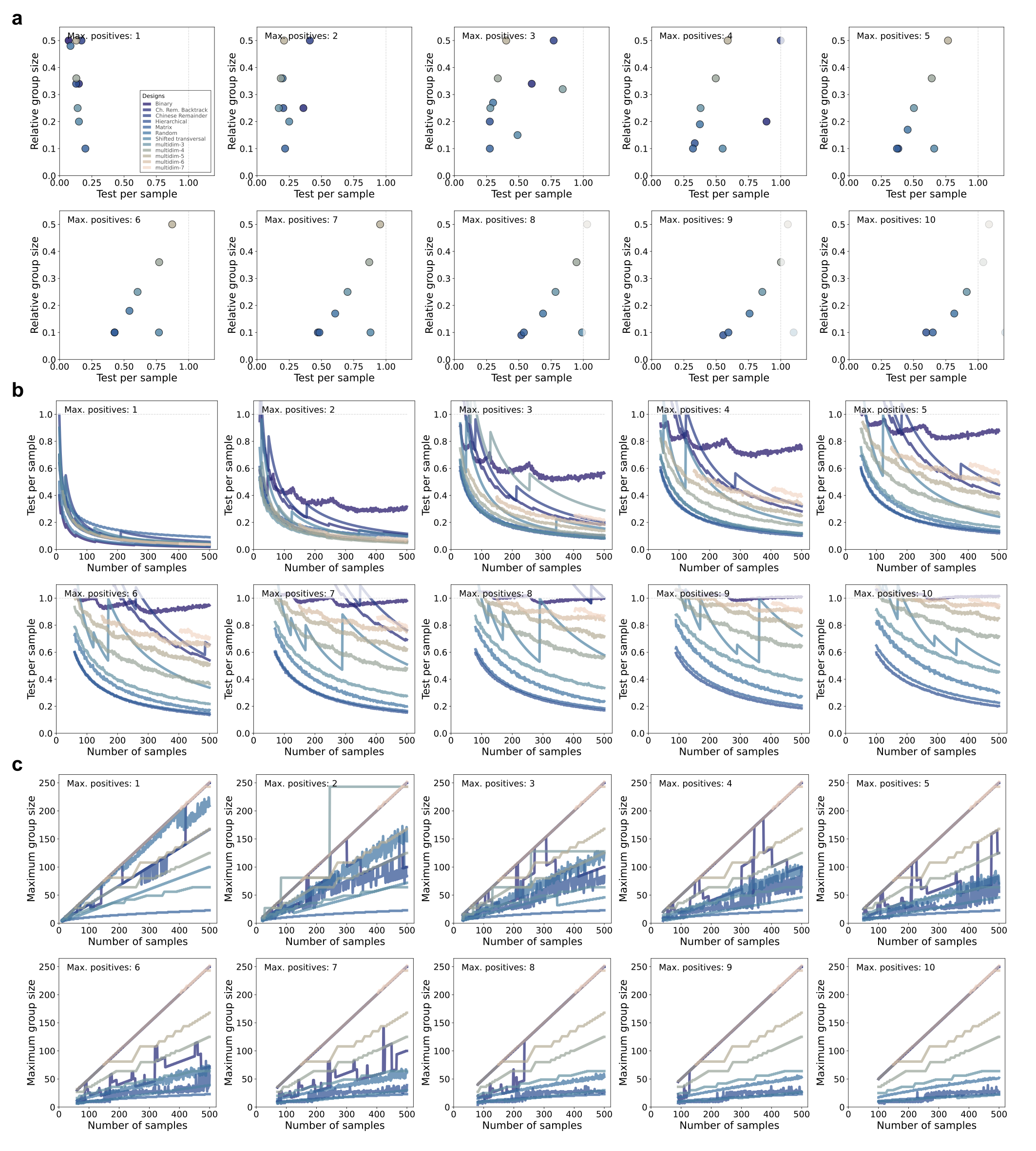}
    \captionsetup{font=footnotesize,singlelinecheck=false}
    \caption{\textbf{Group testing performances and group sizes vary across prevalence values.} (\textbf{a-c}) Relation between relative group size and test number (\textbf{a}), overall numbers of test (\textbf{b}), and maximum group sizes (\textbf{c}) for all 12 PoolPy designs with 1 - 10 maximum number of positive samples across varying numbers of samples. (\textbf{a,b}) The part where group testing becomes less efficient than individually testing each sample (above one test per sample) is grayed out.}
    \label{fig:figSuppGroupSizes}
  \end{minipage}
\end{adjustbox}
\end{figure}
\clearpage

%%%%%%%%%%%%%%%%%%%%%%%%%%%%%%%%%%%%%%%%%%%%%%%%%%%%%%%%%%%%%%%%%%%%%%%%
% FIGURE S3
\begin{figure}[!t]
\centering
\begin{adjustbox}{width=1.35\textwidth,center}
  \begin{minipage}{\linewidth}
    \includegraphics[width=\linewidth]{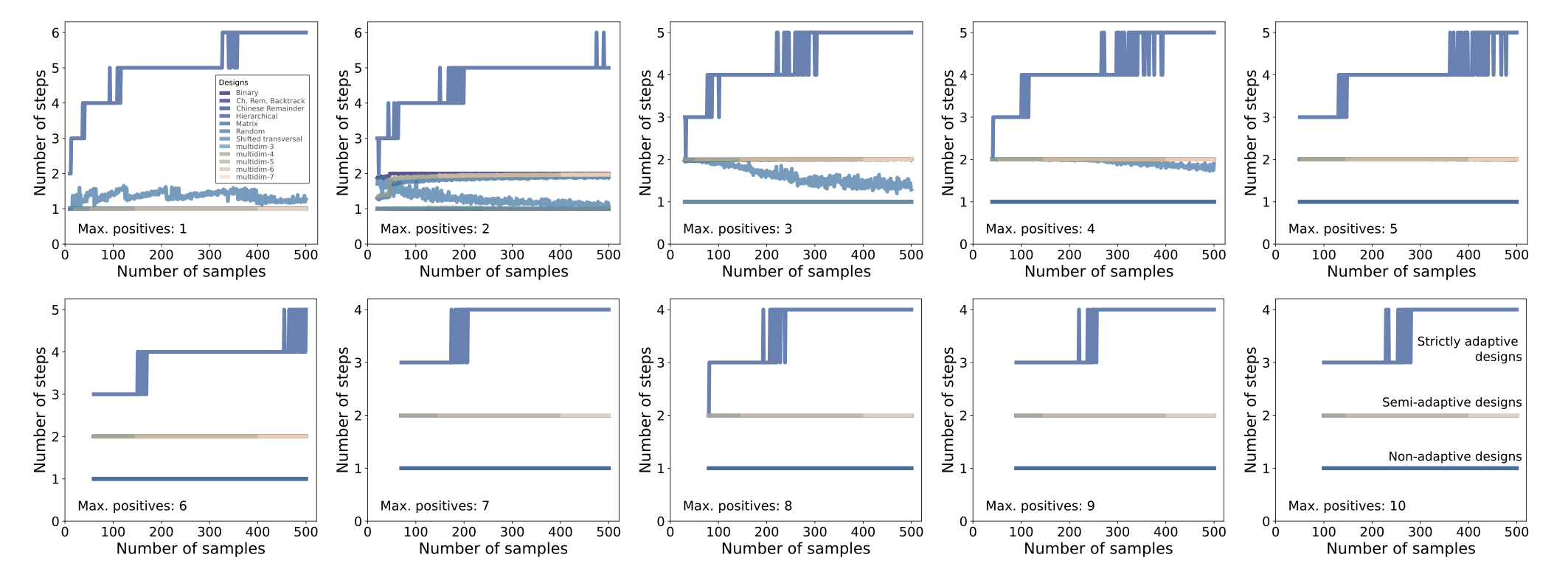}
    \captionsetup{font=footnotesize,singlelinecheck=false}
    \caption{\textbf{Group testing methods require varying numbers of steps.} Number of steps (rounds of experiment) needed using different group testing methods to identify up to 1 - 10 positive samples across varying numbers of samples. Only methods based on the Chinese Remainder Theorem or on the shifted transversal design can identify positive samples in a single step across prevalence values (non-adaptive designs).}
    \label{fig:figSuppSteps}
  \end{minipage}
\end{adjustbox}
\end{figure}
\clearpage

%%%%%%%%%%%%%%%%%%%%%%%%%%%%%%%%%%%%%%%%%%%%%%%%%%%%%%%%%%%%%%%%%%%%%%%%
% FIGURE S4
\begin{figure}[!t]
\centering
\begin{adjustbox}{width=1.35\textwidth,center}
  \begin{minipage}{\linewidth}
    \includegraphics[width=\linewidth]{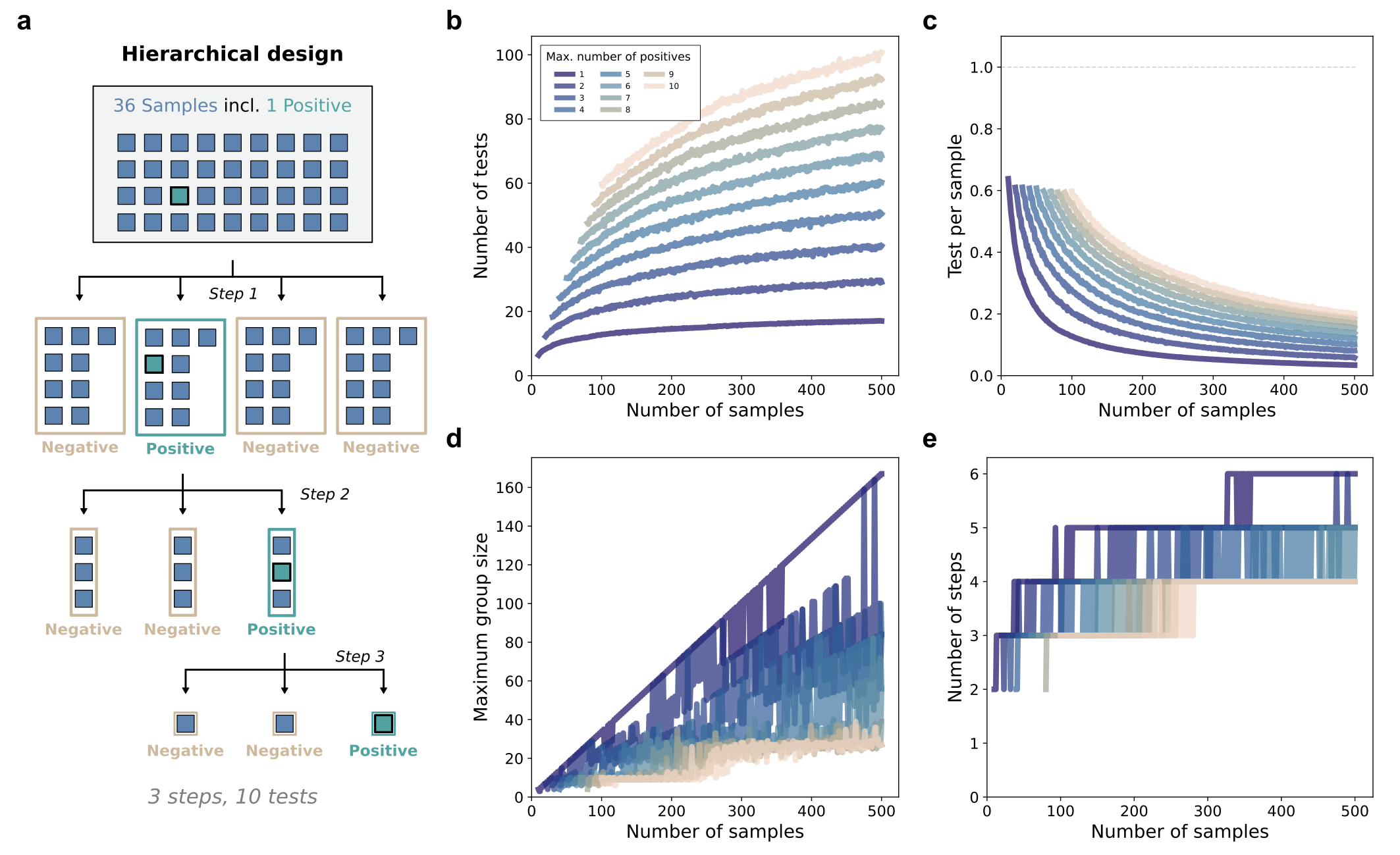}
    \captionsetup{font=footnotesize,singlelinecheck=false}
    \caption{\textbf{The hierarchical method for adaptive group testing.} \textbf{a} Schematic illustration of the hierarchical design. In this example with 36 samples, the hierarchical design uses ten tests across three consecutive steps to identify one positive sample. (\textbf{b-e}) Number of total tests (\textbf{b}), number of test per sample (\textbf{c}), maximum group size (\textbf{d}) or number of steps (\textbf{e}) needed using the hierarchical design with 1 - 10 maximum numbers of positive samples across 10 to 100 samples.}
    \label{fig:figSuppHierarchical}
  \end{minipage}
\end{adjustbox}
\end{figure}
\clearpage

%%%%%%%%%%%%%%%%%%%%%%%%%%%%%%%%%%%%%%%%%%%%%%%%%%%%%%%%%%%%%%%%%%%%%%%%
% FIGURE S5
\begin{figure}[!t]
\centering
\begin{adjustbox}{width=1.35\textwidth,center}
  \begin{minipage}{\linewidth}
    \includegraphics[width=\linewidth]{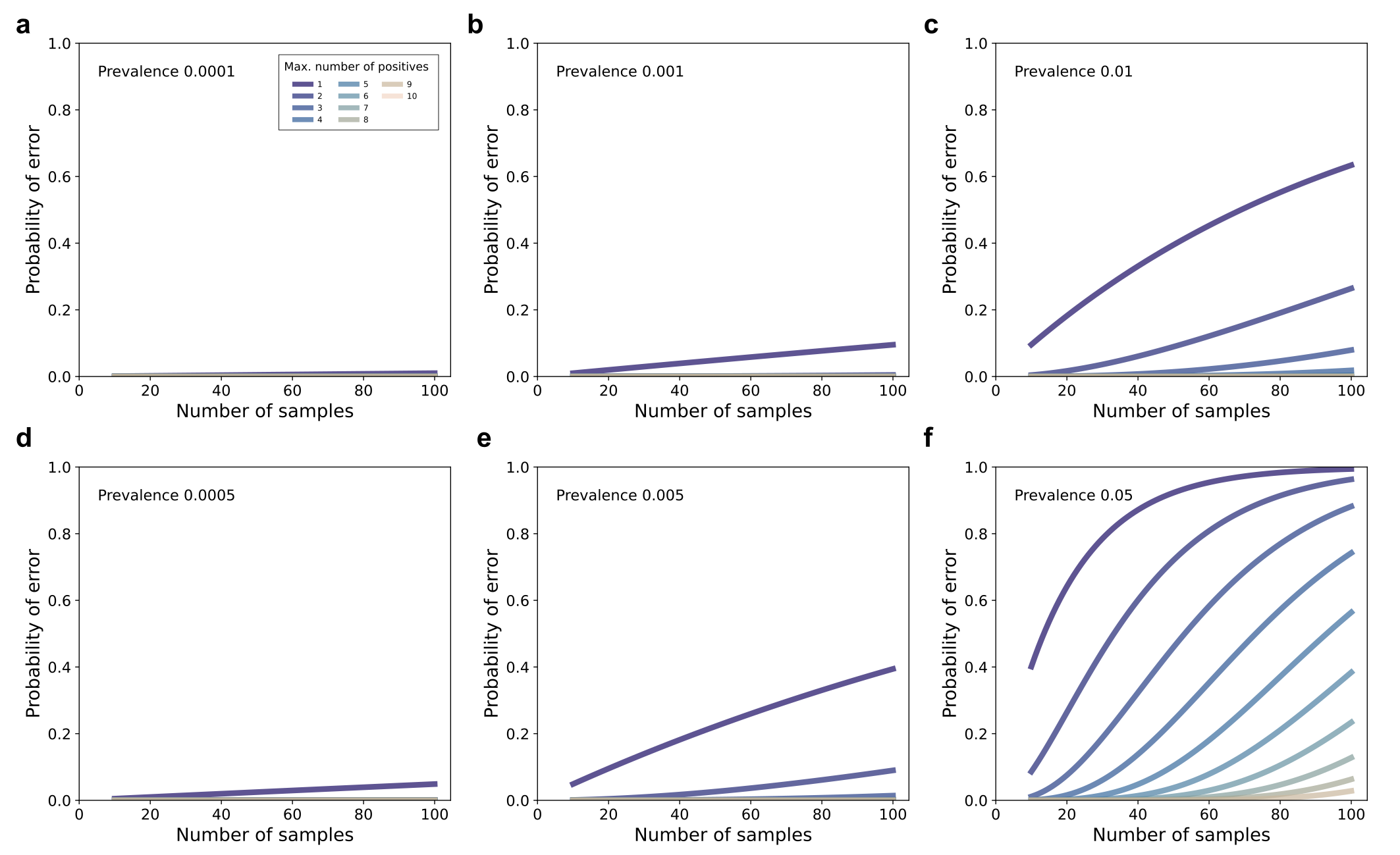}
    \captionsetup{font=footnotesize,singlelinecheck=false}
    \caption{\textbf{Prevalence determines error rates based on expected number of positive samples.} (\textbf{a-f}) Probability of error (identifying the wrong sample(s) as positive) across six prevalence values (\textbf{a} to \textbf{f}) shown for 1 - 10 maximum numbers of expected positive samples.}
    \label{fig:figSuppPrevalence}
  \end{minipage}
\end{adjustbox}
\end{figure}
\clearpage

%%%%%%%%%%%%%%%%%%%%%%%%%%%%%%%%%%%%%%%%%%%%%%%%%%%%%%%%%%%%%%%%%%%%%%%%
% FIGURE S6
\begin{figure}[!t]
\centering
\begin{adjustbox}{width=1.35\textwidth,center}
  \begin{minipage}{\linewidth}
    \includegraphics[width=\linewidth]{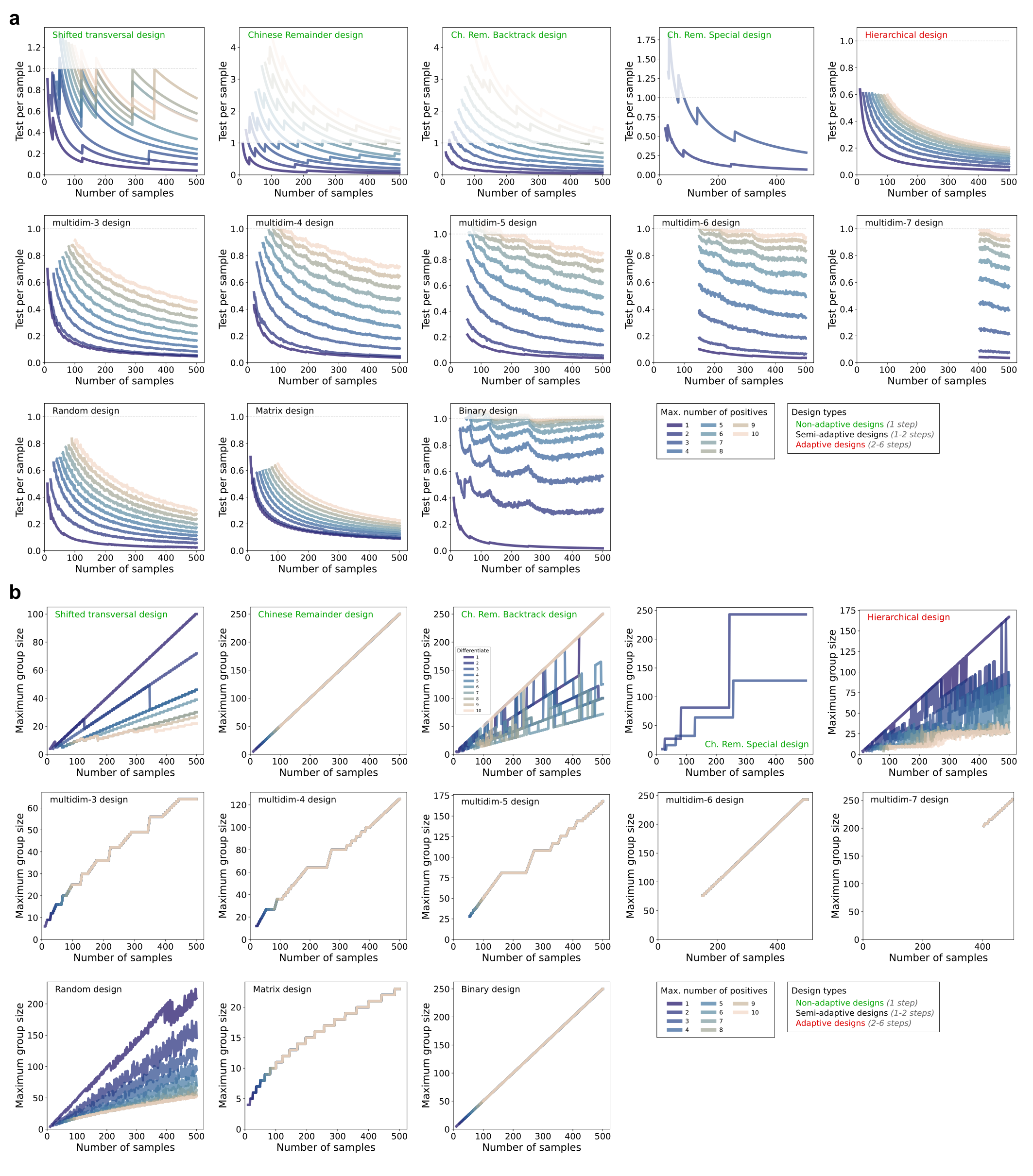}
    \captionsetup{font=footnotesize,singlelinecheck=false}
    \caption{\textbf{Group testing performances across all PoolPy designs.} (\textbf{a-b}) Numbers of test per sample (\textbf{a)} and maximum group sizes (\textbf{b}) for each of the 13 possible PoolPy designs with 1 - 10 maximum numbers of positive samples across 10 to 500 samples.}
    \label{fig:figSuppMethods}
  \end{minipage}
\end{adjustbox}
\end{figure}
\clearpage

%%%%%%%%%%%%%%%%%%%%%%%%%%%%%%%%%%%%%%%%%%%%%%%%%%%%%%%%%%%%%%%%%%%%%%%%
% FIGURE S7
\begin{figure}[!t]
\centering
\begin{adjustbox}{width=1.35\textwidth,center}
  \begin{minipage}{\linewidth}
    \includegraphics[width=\linewidth]{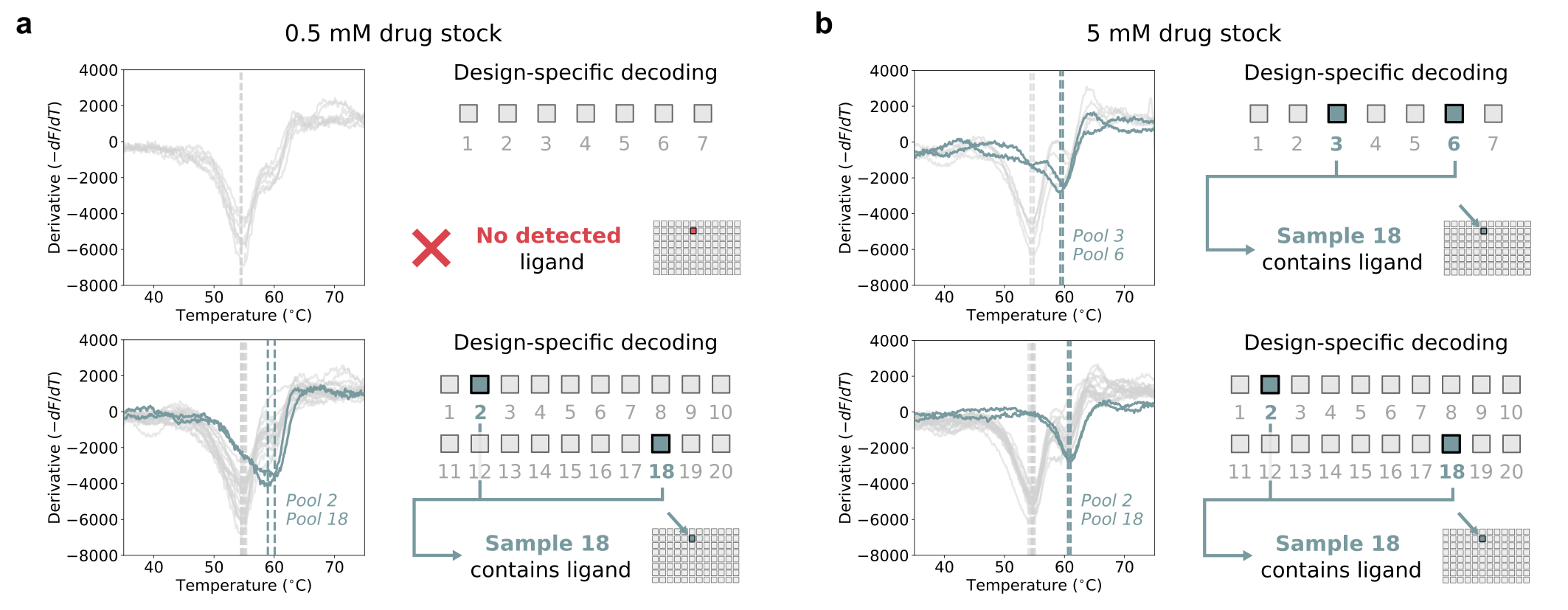}
    \captionsetup{font=footnotesize,singlelinecheck=false}
    \caption{\textbf{Drug stock concentration determines pooling design applicability in protein-ligand screening.} (\textbf{a-b}) Ligand interaction screening for human carbonic anhydrase II of 96 samples including one positive sample containing known ligand acetazolamide (left) and the corresponding decoding scheme (right) for designs minimizing test number (top) or signal dilution (bottom) using acetazolamide (positive) stock concentrations of 0.5mM (\textbf{a}) or 5mM (\textbf{b}).}
    \label{fig:figSuppTSA}
  \end{minipage}
\end{adjustbox}
\end{figure}
\clearpage

%%%%%%%%%%%%
% SUPPLEMENTARY TABLES
% SUPPLEMENTARY TABLES
\renewcommand{\thetable}{S\arabic{table}}
\renewcommand{\arraystretch}{1.5}

%%%%%%%%%%%%%%%%%%%%%%%%%%%%%%%%%%%%%%%%%%%%%%%%%%%%%%%%%%%%%%%%%%%%%%%%
% TABLE S1

\begin{table}[!t]
\centering
\makebox[\linewidth][c]{%
  \begin{adjustbox}{max width=1.35\textwidth}
    \begin{minipage}{\linewidth}

      \captionsetup{font=LARGE,justification=raggedright,singlelinecheck=false}
      
      \rowcolors{2}{white}{background_c}
      \LARGE      
      \begin{tabular}{|c|cccc>{\centering\arraybackslash}m{6cm}c|}
        \toprule
        \textbf{Pooling method}       & \textbf{\begin{tabular}[c]{@{}c@{}}Reducing \\ test number\end{tabular}} 
                              & \textbf{\begin{tabular}[c]{@{}c@{}}Minimizing \\ group size\end{tabular}} 
                              & \textbf{\begin{tabular}[c]{@{}c@{}}Minimizing \\ number of steps\end{tabular}} 
                              & \textbf{\begin{tabular}[c]{@{}c@{}}Scaling with \\ high prevalence\end{tabular}} 
                              & \textbf{Category}              
                              & \textbf{Ref.} \\ 
        \midrule

        \textbf{Hierarchical}          & good        & good        & very poor & very good & Adaptive & \cite{hou2017hierarchical}\\
        \textbf{Binary}                & very good   & very poor   & poor   & very poor & Semi-adaptive & \\
        \textbf{Matrix}                & very poor        & very good   & average   & good   & Semi-adaptive & \cite{du1999combinatorial}\\
       \textbf{Multi-dimensional - 3} & average        & good        & average   & good   & Semi-adaptive & \cite{mutesa2021pooled,barillot1991}\\
        \textbf{Multi-dimensional - 4} & average        & poor     & average   & average      & Semi-adaptive & \cite{mutesa2021pooled,barillot1991}\\
        \textbf{Multi-dimensional - 5} & average        & very poor     & average   & average      & Semi-adaptive & \cite{mutesa2021pooled,barillot1991}\\
        \textbf{Random}                & good        & average     & poor   & good      & Semi-adaptive & \cite{bruno1995efficient}\\
        \textbf{Shifted transversal}   & poor        & good        & very good & average      & Non-adaptive & \cite{thierry2006new,xin2009shifted}\\
        \textbf{Chinese Remainder}     & very poor        & very poor   & very good & very poor      & Non-adaptive & \cite{eppstein2007improved}\\
        \textbf{Ch. Rem. Backtrack}    & poor     & average     & very good & poor      & Non-adaptive & \cite{eppstein2007improved}\\
        \textbf{Ch. Rem. Special}      & poor        & poor   & very good & poor      & Non-adaptive & \cite{eppstein2007improved}\\

        \bottomrule
      \end{tabular}
      \captionsetup{font=LARGE,justification=justified,singlelinecheck=false}
      \caption{\textbf{Performance comparison of group testing methods supported by PoolPy.} Methods were ranked based on four criteria reflecting real-world constraints to guide user choice. The methods were classified in quintile ranks (corresponding to the five annotations in order: very poor, poor, average, good and very good) for their average value of the corresponding metric across all designs from sample set sizes of 20 to 100. For test number and group size, designs made with a maximum number of positives of one were used. For number of steps and scaling at high prevalence, the average over all designs made with maximum number of positives values of 1 - 5 was used.}
      \label{tab:methods}
    \end{minipage}
  \end{adjustbox}%
}
\end{table}
\renewcommand{\arraystretch}{1.0}
\clearpage

%%%%%%%%%%%%%%%%%%%%%%%%%%%%%%%%%%%%%%%%%%%%%%%%%%%%%%%%%%%%%%%%%%%%%%%%
%%%%%%%%%%%%%%%%%%%%%%%%%%%%%%%%%%%%%%%%%%%%%%%%%%%%%%%%%%%%%%%%%%%%%%%%
% TABLE S2 PROVIDED AS FILE
%%%%%%%%%%%%%%%%%%%%%%%%%%%%%%%%%%%%%%%%%%%%%%%%%%%%%%%%%%%%%%%%%%%%%%%%
\addtocounter{table}{1}

%%%%%%%%%%%%%%%%%%%%%%%%%%%%%%%%%%%%%%%%%%%%%%%%%%%%%%%%%%%%%%%%%%%%%%%%
% TABLE S3

\begin{table}[!t]
\centering
\makebox[\linewidth][c]{%
  \begin{adjustbox}{max width=1\textwidth}
    \begin{minipage}{\linewidth}

      % Match the caption style of the original
      \captionsetup{font=LARGE,justification=raggedright,singlelinecheck=false}
      
      % Alternating row colors (requires xcolor with [table] option)
      \rowcolors{2}{white}{background_c}
      \setlength{\extrarowheight}{6pt}
      \LARGE      
      
      \begin{tabular}{|cccccc|} % Removed vertical bars for the "booktabs" look
        \toprule
        \textbf{TF ID} & \textbf{TF name} & \textbf{Gene ID} & \textbf{Known binding sites} & \textbf{Recovered single} & \textbf{Recovered pooled}\\ 
        \midrule
        A & PdhR & b0113 & 15 & 7 & 10\\ 
        B & TorR & b0995 & 11 & 4 & 4\\ 
        C & RutR & b1013 & 5  & 3 & 3\\ 
        D & DhaR & b1201 & 0  & - & -\\ 
        E & Mlc  & b1594 & 7 & 7 & 6 \\ 
        F & GlpR & b3423 & 18 & 2 & 2 \\ 
        G & MetJ & b3938 & 27 & 19 & 15\\ 
        H & IclR & b4018 & 8 & 5 & 8 \\ 
        I & UxuR & b4324 & 8 & 5 & 5 \\ 
        J & DgoR & b4479 & 2 & 2 & 0 \\ 
        \bottomrule
      \end{tabular}

      \captionsetup{font=LARGE,justification=justified,singlelinecheck=false}
      \caption{\textbf{Transcription factors binding site summary.} This table lists the ten \textit{E. coli} TFs studied in this work using either single or pooled DAP-seq. The numbers of previously known binding sites as well as of recovered binding sites using either single or pooled assays are indicated.}
      \label{tab:TFTable}
    \end{minipage}
  \end{adjustbox}%
}
\end{table}
\renewcommand{\arraystretch}{1.0}
\clearpage

%%%%%%%%%%%%%%%%%%%%%%%%%%%%%%%%%%%%%%%%%%%%%%%%%%%%%%%%%%%%%%%%%%%%%%%%
% TABLE S4

\begin{table}[!t]
\centering
\makebox[\linewidth][c]{%
  \begin{adjustbox}{max width=1.4\textwidth}
    \begin{minipage}{\linewidth}

      \captionsetup{font=large,justification=raggedright,singlelinecheck=false}
      
      \rowcolors{2}{white}{background_c}
      \setlength{\extrarowheight}{6pt}
      \large      
      
      % Using 'p' column for sequence to allow wrapping of long DNA strings
      \begin{tabular}{|l p{20cm} l|} 
        \toprule
        \textbf{Name} & \textbf{Sequence} & \textbf{Notes} \\ 
        \midrule
        Adaptor A    & CACGACGCTCTTCCGATCT & \\ 
        Adaptor B    & P-GATCGGAAGAGCACACGTCTG & "P-": phosphorylation \\ 
        P1           & CACGACGCTCTTCCGATCT & \\ 
        P2           & CAGACGTGTGCTCTTCCGATC & \\ 
        P5\_universal & AATGATACGGCGACCACCGAGATCTACACTCTTTCCCTACACGACGCTCTTCCGATCT & \\ 
        P7\_barcoded  & CAAGCAGAAGACGGCATACGAGATXXXXXXXXGTGACTGGAGTTCAGACGTGTGCTCTTCCGATC & "XXXXXXXX": barcode \\ 
        \bottomrule
      \end{tabular}

      \captionsetup{font=large,justification=justified,singlelinecheck=false}
      \caption{\textbf{Primers used in this study.} List of adaptors and primers used in the DAP-seq protocol. Sequences are reported in the 5' to 3' direction.}
      \label{tab:primers}
    \end{minipage}
  \end{adjustbox}%
}
\end{table}
\renewcommand{\arraystretch}{1.0}
\clearpage

\newpage
%%%%%%%%%%%%%%%%%%%%%%%%%%%%%%%%%%%%%%%%%%%%%%%%%%%%%%%%%%%%%%%%%%%%%%%%%%%%%%%%%%%%%%%%%%%%%
\subsection*{Supplementary Note 1. The foundational group testing designs.}\label{supp_note_1}
% Note Fig 1
\begin{wrapfigure}{r}{8cm}
    \includegraphics[width=8cm]{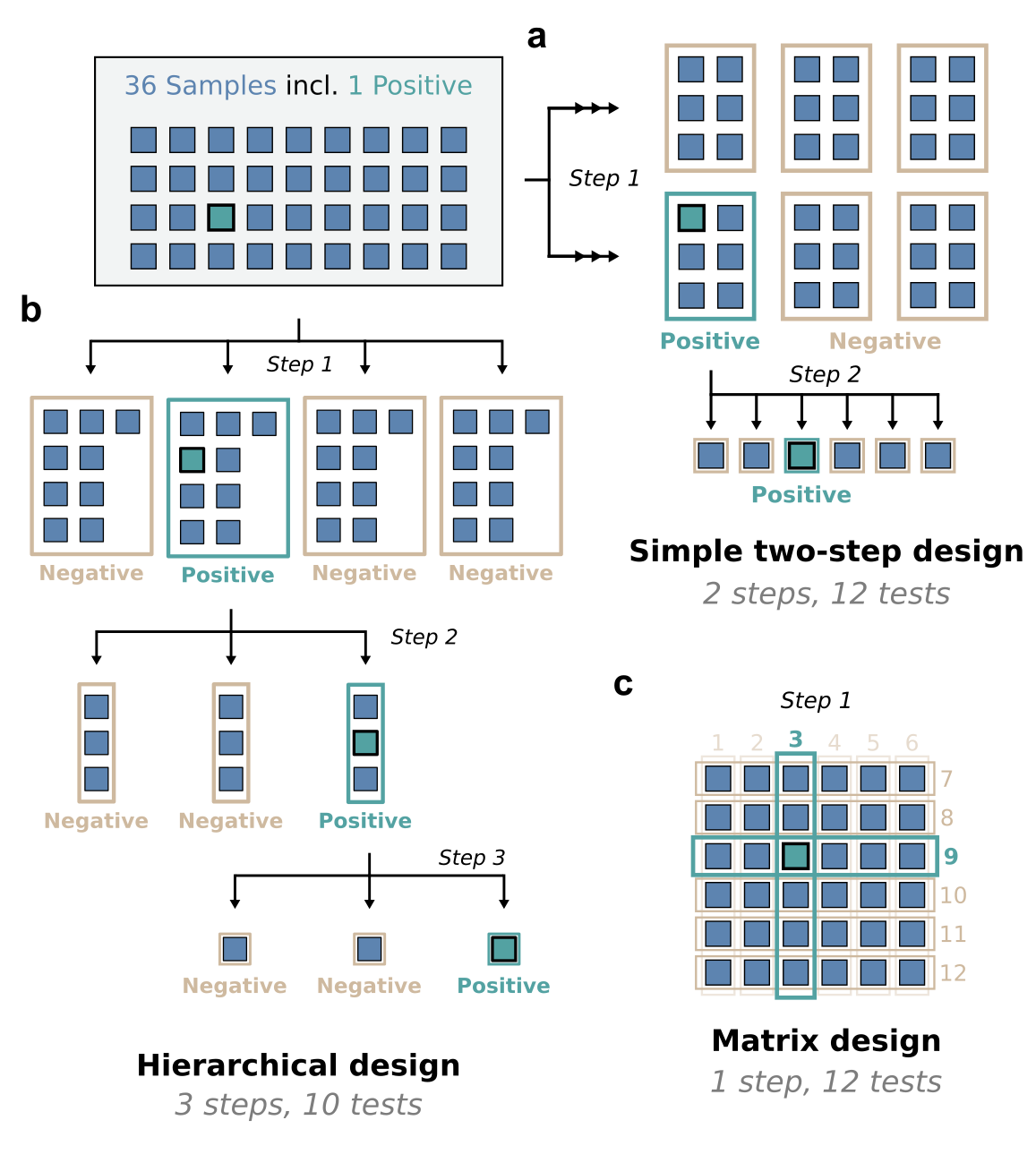}
    \captionsetup{font=footnotesize,singlelinecheck=false}
    \caption*{\textbf{Schematic illustration of simple pooling designs.} In this example, one positive sample has to be identified out of 36 (top left). Three possible designs are illustrated, using the simple two-step (\textbf{a}), hierarchical (\textbf{b}) or matrix (\textbf{c}) designs.}
    \label{fig:SuppNote_Fig_fundational}
\end{wrapfigure}

Group testing started with simple, intuitive designs (\hyperref[fig:fig1]{Fig. 1a}). It was first introduced during World War II as a way to reduce the number of tests needed to identify soldiers with syphilis infections \cite{dorfman1943detection}. There, a simple two-step design was used where samples were pooled and tested in small groups, and when a group came back positive, each sample in it were tested individually in a second step (\textbf{a}). An improvement of this method in terms of number of tests was made in the hierarchical design \cite{hou2017hierarchical}, where a more progressive step-by-step approach is used (\textbf{b}). While both designs reduce the number of tests significantly, they require multiple steps, making them less attractive when quick results are needed.
The Matrix design offers another intuitive approach by arranging samples in a two-dimensional grid \cite{du1999combinatorial}. Each row and each column of the grid is pooled and tested as a group, so that a positive result in a specific row and column identifies the corresponding positive sample in a single step (\textbf{c}). However, this design is particularly sensitive to increase in prevalence, as additional steps are necessary if more than one positive sample is present in the tested set.
These designs illustrate a common trade-off in group testing: fewer testing steps, as in the matrix design, usually come at the cost of flexibility when multiple positives are present, while multi-step methods, such as the hierarchical design, handle several positives more reliably but require more rounds of testing. Some modern non-adaptive methods, however, can resolve multiple positive samples in one step efficiently across prevalence values, partially bridging this gap.

%%%%%%%%%%%%%%%%%%%%%%%%%%%%%%%%%%%%%%%%%%%%%%%%%%%%%%%%%%%%%%%%%%%%%%%%%%%%%%%%%%%%%%%%%%%%%
\newpage
\subsection*{Supplementary Note 2. Methods implemented in PoolPy.}\label{supp_note_2}

PoolPy supports nine conceptually-different group testing algorithms, which can be classified based on two properties: (i) whether their designs depend on a maximum number of positives (differentiate-dependent methods) or not (differentiate-independent methods), and (ii) whether they can identify positive samples in a single step consistently (non-adaptive methods), in some cases (semi-adaptive methods), or always require multiple steps (adaptive methods). The supported methods are as follows:

\subsubsection*{Differentiate-dependent methods}
These methods adapt their designs across expected prevalence values, even with fixed numbers of samples, enabling them to scale better with larger numbers of positive samples.
\begin{itemize}
  \item \textbf{Random}: A semi-adaptive method that pools samples at random according to user-defined constraints. This method might require a second step of validation testing to resolve ambiguities.
  \item \textbf{Shifted Transversal Design} (STD): A non-adaptive method based on prime numbers and modular arithmetic. Each sample is assigned to pools according to shifted transversal patterns that guarantee unique identification up to a maximum number of positive samples. This method can consistently identify positive sample(s) in a single step when prevalence assumptions are met.
  \item \textbf{Chinese Remainder methods}: Three non-adaptive methods based on the Chinese Remainder Theorem which encode sample identities using modular congruences across multiple pools. PoolPy supports a standard variant, a backtracking variant (backtrack) that resolves ambigous cases by exploring compatible solutions and a special variant (special) optimized for 2 and 3 maximum number of positive samples. These methods can consistently identify positive sample(s) in a single step when prevalence assumptions are met.
  \item \textbf{Hierachical}: A strictly adaptive method that iteratively splits the set of possible positive samples. Positive pools are subdivided in subsequent steps until positive samples are identified. This method consistently requires multiple steps.
\end{itemize}

\subsubsection*{Differentiate-independent methods}
These methods generate designs that do not depend on an expected prevalence value. They can show very good performances at low prevalence, but should be used with caution if multiple positive samples are expected.
\begin{itemize}
  \item \textbf{Matrix}: A semi-adaptive method where samples are arranged in a two-dimensional grid, with each row and each column forming a pool. This method might require a second step of validation testing to resolve ambiguities with more than a single positive sample.
  \item \textbf{Multidimensional methods}: A set of semi-adaptive methods where samples are arranged in higher-dimensional matrices (3D, 4D, 5D, 6D and 7D are implemented in PoolPy), where each coordinate along each dimension corresponds to a pool. These methods might require a second step of validation testing to resolve ambiguities with more than a single positive sample.
  \item \textbf{Binary}: A semi-adaptive method where samples are assigned to pools according to a binary code, maximizing information per test. This method might require a second step of validation testing to resolve ambiguities with more than a single positive sample.
\end{itemize}

%%%%%%%%%%%%%%%%%%%%%%%%%%%%%%%%%%%%%%%%%%%%%%%%%%%%%%%%%%%%%%%%%%%%%%%%%%%%%%%%%%%%%%%%%%%%%
\newpage
\subsection*{Supplementary Note 3. The logistical constraints of group testing.}\label{supp_note_3}

For simplicity, we refer to samples exhibiting the particular property detected by the test used as \textit{positive}, and to their proportion within the entire set of possible samples as the \textit{prevalence}. 
Across biomedical fields, various terms may be used for these two concepts; they are interchangeable and fully compatible with PoolPy.

\textbf{Prevalence.} The performance of combinatorial group testing  designs is highly affected by prevalence, such as the proportion of infected individuals in a population, the hit rate in a drug screening, or the likelihood of detecting a specific feature in molecular profiling. At very low prevalence, group testing can greatly reduce the number of tests, while higher prevalence quickly reduces its efficiency \cite{aldridge2019group}. Using sample sets that contain more positive samples than a set threshold (the maximum number of positives) will typically lead to inconclusive results. Thus, it is important to obtain an estimate of the prevalence in the tested population \textit{a priori} to choose an appropriate design. In its Prevalence section, PoolPy offers a tool to estimate error rates and choose appropriate threshold values and pooling parameters to match the expected prevalence. Differentiate-dependent methods adapt their design based on the expected maximum number of positive samples, and typically scale well with increased prevalence. Conversely, while differentiate-independent methods show attractive performance with at most one positive sample, their performance drastically decreases with more positives, often leading to inconclusive results in these cases.\\

Combinatorial group testing tends to lose its efficacy at higher prevalence.
Indeed, for any given prevalence $\rho$ we can have a lower bound on the information in bits encoded in a system with $N$ samples:
\begin{equation}
    I\ge\binom{N}{\rho N}=\frac{N!}{(\rho N)!((1-\rho)N)!}.
\end{equation}
This value is a lower bound as it assumes that exactly $\rho N$ samples are positive, while in experimental settings this is generally unknown.
Using Stirling's approximation
\begin{equation}
    N!\sim \sqrt{2\pi N} \left( \frac{N}{e}\right)^N
\end{equation}
we can write assuming $\rho N \in \mathbb{N}$:
\begin{equation}
    I(\rho)\ge \frac{ \sqrt{2\pi N}\left( \frac{N}{e}\right)^N}{\sqrt{2\pi (\rho N)} \left( \frac{(\rho N)}{e}\right)^{(\rho N)}  \sqrt{2\pi ((1-\rho)N)}\left( \frac{((1-\rho)N)}{e}\right)^{((1-\rho)N)}}.
\end{equation}
This expression recapitulates the limit of low prevalence $\rho=1/N$ with $I(\rho)\ge N$ where we find methods that need $\ceil{\text{log}_2(N)}$ tests.
Using this formula, we can estimate the upper limit of prevalence, $\rho=1/2$.
\begin{equation}
    I(1/2)\ge \frac{ \left( \frac{N}{e}\right)^N}{\left( \frac{(N/2)}{e}\right)^{N} }=2^N.
\end{equation}
In this case, at least $\ceil{\text{log}_2(2^N)}=N$ tests are needed, rendering the combinatorial pooling approach futile.
It is important to note that this information limit is reached for practical purposes before $\rho=1/2$ prevalence.

\textbf{Turnaround time.} Adaptive group testing designs require multiple steps, such as the hierarchical method \cite{hou2017hierarchical}, where the outcome of one round of testing determines how the next round has to be set up. While these designs often require low overall number of tests, the total time until results are available is increased due to the multiple steps. Semi-adaptive designs can identify positive samples in a single step with low prevalence but often require a second validation step if more than one positive sample is present. Lastly, non-adaptive designs can consistently identify positive samples in a single step. All non-adaptive designs implemented in PoolPy rely on a differentiate-dependent method to allow design adaptation to higher prevalence values.
In use cases where fast turnaround time is essential, non-adaptive methods that always require a single step, such as the chinese remainder or the shifted transversal methods \cite{eppstein2007improved,xin2009shifted}, may be preferable despite the fact that they may require more tests than adaptive designs.\\
\\[0.1em]
\textbf{Group size limit.} Practical limits exist on how many individual samples can be pooled without loss of sensitivity, which mostly depend on the nature of the test being used. This is often referred to as the \textit{signal dilution effect} in biomedical applications where the use of very large pools can increase the risk of false negatives due to increased dilution. Depending on applications, a group size limit can be defined to guide the choice of appropriate group testing designs, as some designs use relatively large group sizes. This is entirely application-specific and depends on assay sensitivity.\\

With increased relative group sizes, the number of times that each sample is tested can increase as well. This represents a second parameter depending on group definition that can affect applicability. In cases where the amount of sample is limiting, or for valuable samples, users should choose designs that use small group sizes to reduce sample usage.\\
\\[0.1em]
\textbf{Assay type.} PoolPy supports any assay that is compatible with group testing, which can typically be described as single- or multi-readout. Single-readout assays have a single binary outcome, such as infection testing (positive/negative) or ligand-target interaction screening (interacting/not interacting). Multi-readout assays yield multiple outcomes also classifiable in a binary manner. They include complex molecular profiling methods relying on omics measurements where each measured unit (e.g. each gene, protein, binding event ...) is considered as an individual outcome. In the case of multi-readout assays, the choice of pooling design applies to all outcomes, and should thus be selected based on the most stringent pooling parameters out of all possible outcomes.

%%%%%%%%%%%%%%%%%%%%%%%%%%%%%%%%%%%%%%%%%%%%%%%%%%%%%%%%%%%%%%%%%%%%%%%%%%%%%%%%%%%%%%%%%%%%%
\newpage
\subsection*{Supplementary Note 4. PoolPy user guide.}\label{supp_note_4}

PoolPy is provided as a web interface (\href{https://trouillon-lab.github.io/PoolPy/}{https://trouillon-lab.github.io/PoolPy}) where users can design and compare group testing strategies under defined constraints. The web interface comprises five sections with distinct functions, as follows:
\begin{itemize}
  \item \textbf{Methods Comparison}: The main section of PoolPy, where users can obtain direct comparisons of methods performance and download group testing designs for their applications based on a number of samples and a maximum number of positives.
  \item \textbf{Decoder}: This section provides a tool for decoding results from a group testing experiment. It takes as input the design used and the list of positive groups. The decoder also works for multi-readout assays when providing a corresponding readout file.
  \item \textbf{Prevalence}: This section provides a tool for selecting design parameters, such as an expected maximum number of positives and a number of batches, based on the estimated prevalence and an error rate threshold.
  \item \textbf{Automation}: This section provides a tool for generating method files for liquid-handling robots to follow the pooling scheme of a group testing design.
  \item \textbf{Help}: This section describes the methods and metrics used in the other sections.
\end{itemize}

It is also possible to use PoolPy locally without using the web interface, following the instructions in the PoolPy GitHub repository (\href{https://github.com/trouillon-lab/PoolPy}{https://github.com/trouillon-lab/PoolPy}). Local use might be required in cases that are computationally intensive, such as to generate a design with the random method or to get a full performance summary table for cases that were not precomputed. However, the vast majority of use cases are covered by the web interface, which has precomputed designs across \textgreater100,000 conditions, and can generate design on the fly for any numbers of samples and positives for 11 out of 12 methods. Here, we provide a step-by-step guide on how to use the different sections of the PoolPy web interface to perform a group testing experiment. 

PoolPy supports both single-readout assays with binary outcomes, such as diagnostic infection testing or ligand–target interaction screening, and multi-readout experiments, including mass spectrometry- or sequencing-based molecular profiling approaches. In both cases, the samples can be pooled following the generated PoolPy designs. For multi-readout assays, the highest expected prevalence value should be used to choose pooling parameters.

\subsubsection*{1. Determine pooling parameters based on prevalence (optional)}

\begin{figure}[t]
\centering
    \includegraphics[width=\linewidth]{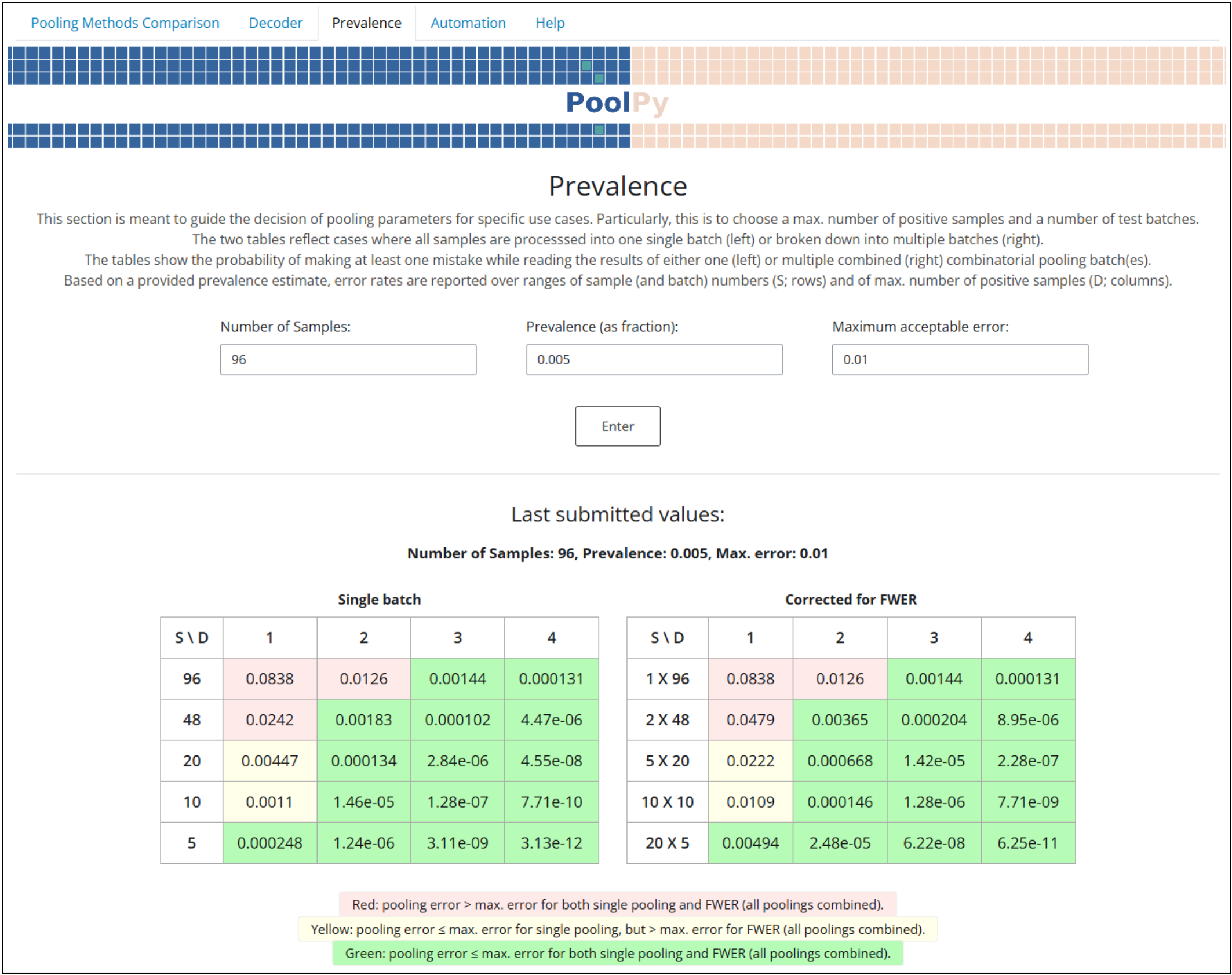}
    \captionsetup{font=footnotesize,singlelinecheck=false}
    \caption*{\textbf{Screenshot 1: Pooling parameter determination using the PoolPy Prevalence section.} }
    \label{fig:SuppNote_Fig1}
\end{figure}

Since many group testing methods adapt their design to an expected number of positive samples, it is important to estimate the prevalence, or hit rate, among the tested samples before choosing a design. This estimated prevalence can then be used to define the pooling parameters. This can be done in the Prevalence section of the PoolPy web interface (\hyperref[fig:SuppNote_Fig1]{Screenshot 1}). There, users can input their test conditions -- the number of samples, the estimated prevalence, and the maximum acceptable error -- to get the probability of making a potential mistake when reading the results of a pooled experiment; i.e. when more positive samples are present than the defined maximum number of positives. The consequences of such a mistake differ between methods and pooling parameters, ranging from needing to perform a validation round of a few more tests to getting completely inconclusive results, and thus should ideally be avoided.

In the example shown in \hyperref[fig:SuppNote_Fig1]{Screenshot 1}, the probability of error is shown for a case with 96 samples with an estimated prevalence of 0.5\% across different numbers of batches and maximum numbers of positive samples (\textit{differentiate}). To keep the probability of making at least one mistake below 1\%, several strategies are possible in this example, such as performing a pooling experiment with all samples and a \textit{differentiate} value of 3, or splitting the samples into two batches of 48 samples each and a \textit{differentiate} value of 2. With this overview, users can decide on viable combinations of batch number and \textit{differentiate} values to use to design their pooling experiment in the PoolPy Methods Comparison section. 

% Screenshot 1

\subsubsection*{2. Compare and choose group testing design}

With a defined number of samples and maximum number of positive samples, users can compare the methods supported by PoolPy across different key performance indicators using the Methods Comparison section. In this section, precomputed performance summary tables are available for a large range of numbers of samples. In case a precomputed performance summary table is not available, specific designs can still be generated and downloaded for 11 out of 12 methods for any number of samples and positives.

% Screenshot 2
\begin{figure}[t]
\centering
    \includegraphics[width=\linewidth]{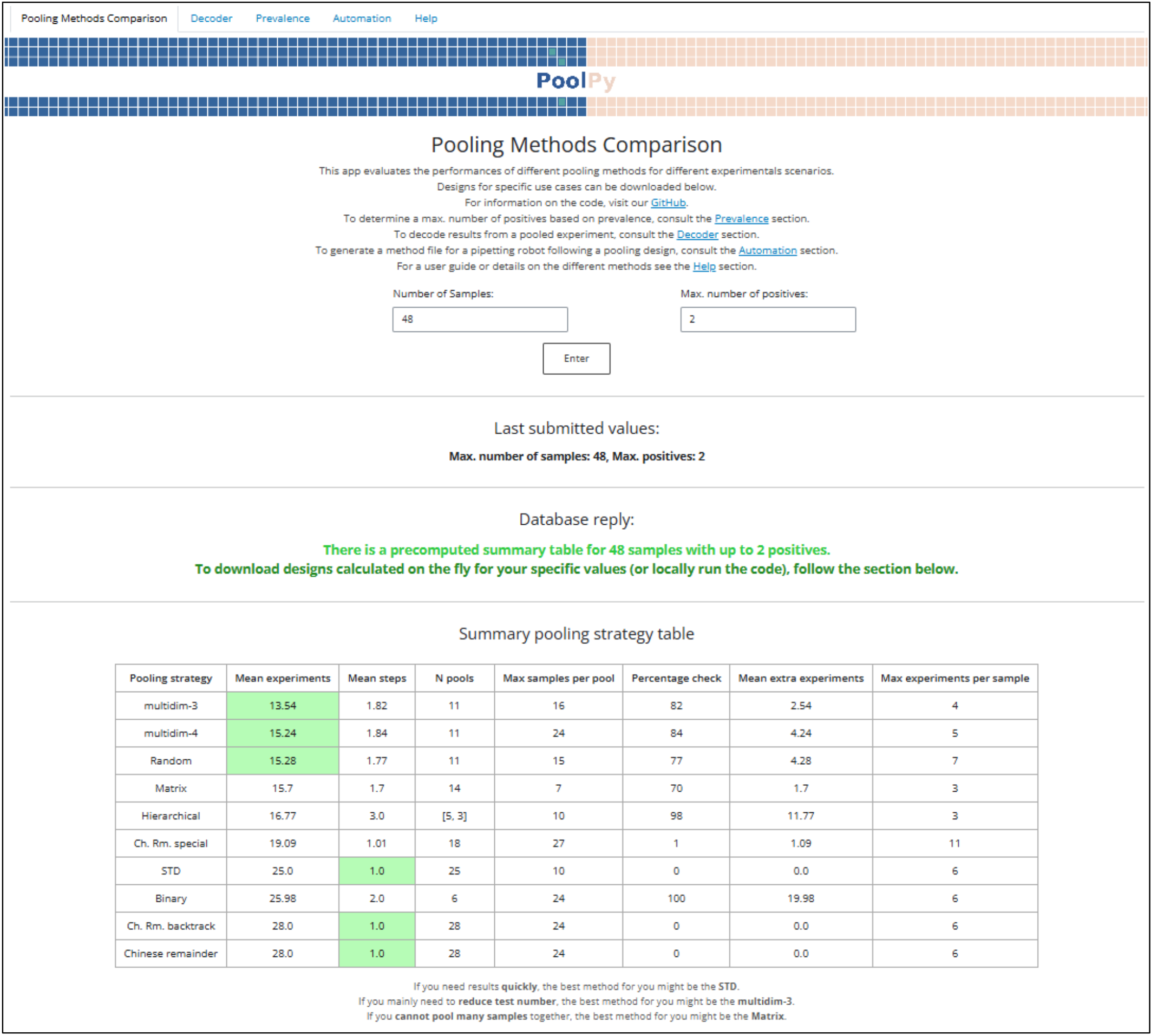}
    \captionsetup{font=footnotesize,singlelinecheck=false}
    \caption*{\textbf{Screenshot 2: Comparing and selecting group testing design using PoolPy.} }
    \label{fig:SuppNote_Fig2}
\end{figure}

Following the suggestions from the Prevalence section above, we show an example of a method comparison for 48 samples with a \textit{differentiate} value of 2 given by PoolPy (\hyperref[fig:SuppNote_Fig2]{Screenshot 2}). In these conditions, different designs are suggested based on different prioritized constraints. To minimize turnaround time, test number or sample dilution, the shifted transversal, the multidim-3 and the matrix designs are suggested, respectively. A summary table is given that details all values for these key metric. Based on these values and suggestions, users can decide on a design of choice and can directly download the corresponding design file from that section. The design file is a table indicating in which pool(s) each sample has to be added.

\subsubsection*{3. Generate automation file for pipetting robot (optional)}

Although the design files can be used directly from the previous section for manual pooling of samples, PoolPy also has an Automation section allowing users to generate method files for pipetting robots to automatize the pooling of samples. In this section, users can input a design file obtained from the Methods Comparison section (or from the PoolPy python code) and directly download method files for pipetting robots from several of the main manufacturers. The downloaded method file can then be used directly on the corresponding robot to perform the pooling of the samples following the selected design.

\subsubsection*{4. Decode pooling experiment results}

After performing the sample pooling and corresponding experiment, users can decode the results of their tests on the Decoder section of PoolPy. There, users can input their design file obtained from the Methods Comparison section (or from the PoolPy python code) and write their readout, i.e. which pools were positive (or any other binary characteristic of interest) in their experiment. The decoder then outputs the identity of the positive sample(s) based on the pooled results. In a successful pooling experiment, the decoder will identify the exact set of positive samples. In some cases, e.g. if the number of positive samples exceeds the defined maximum number of positives set by the user, the decoder might not be able to identify the exact set of positive samples, but will rather output a list of possible samples that contains the positive ones.

For multi-readout assays, users can upload a readout file that contains the list positive pools for each individual readout (one per line). PoolPy will then decode each readout and output a result file with the identity of the positive samples for each specified readout.

\end{document}